\pdfoutput=1
\documentclass[12pt]{article}
\usepackage{multirow}
\usepackage{amsthm,amsmath,amsfonts,natbib}
\usepackage{verbatim}
\usepackage{graphicx}
\usepackage{caption}
\usepackage{tabu}
\usepackage{booktabs}
\usepackage[font=footnotesize]{caption}
\usepackage{subcaption}
\usepackage{bm}
\usepackage{bbm}
\usepackage{xcolor}
\usepackage{bbm}
\usepackage{float}
\floatstyle{plaintop}
\restylefloat{table}
\RequirePackage[colorlinks,citecolor=blue,urlcolor=blue]{hyperref}
\usepackage{natbib}
\usepackage{url} 

\newcommand{\blind}{1}

\begin{document}

\def\spacingset#1{\renewcommand{\baselinestretch}%
{#1}\small\normalsize} \spacingset{1}


\if1\blind
{
 \title{\bf A changepoint approach for the identification of financial extreme regimes}
  \author{Chiara Lattanzi \thanks{
    Part of this work was carried out during CL final year project of the double-degree program in Statistics between the University of Bologna and the University of Glasgow.}\hspace{.2cm}\\
    Dipartimento di Scienze Statistiche, Universit\'{a} di Bologna\\
    and \\
    Manuele Leonelli \\
    School of Mathematics and Statistics, University of Glasgow}
  \maketitle
} \fi

\if0\blind
{
  \bigskip
  \bigskip
  \bigskip
  \begin{center}
    {\LARGE\bf Title}
\end{center}
  \medskip
} \fi

\bigskip
\begin{abstract}
Inference over tails is usually performed by fitting an appropriate limiting distribution over observations that exceed a fixed threshold. However, the choice of such threshold is critical and can affect the inferential results. Extreme value mixture models have been defined to estimate the threshold using the full dataset and to give accurate tail estimates. Such models assume that the tail behavior is constant for all observations. However, the extreme behavior of financial returns  often changes considerably in time and such changes occur by sudden shocks of the market. Here we extend the extreme value mixture model class to formally take into account distributional extreme changepoints, by allowing for the presence of regime-dependent parameters modelling the tail of the distribution.  This extension formally uses the full dataset to both estimate the thresholds and the extreme changepoint locations, giving uncertainty measures for both quantities. Estimation of functions of interest in extreme value analyses is performed via MCMC algorithms. Our approach is evaluated through a series of simulations, applied to real data sets and assessed against competing approaches. Evidence demonstrates that the inclusion of different extreme regimes outperforms both static and dynamic competing approaches in financial applications.
\end{abstract}

\noindent%
{\it Keywords:} 
Extreme value mixture models; Financial returns; GPD distribution; High quantiles; Threshold estimation.

\spacingset{1.45} 
\section{Introduction}
\label{sec:intro}
The financial market is characterized by periods of turbolence where extreme events shock the system, potentially leading to huge profit losses. For this reason it is fundamental to understand and predict the tail distribution of financial returns. As claimed in \citet{Rocco2014}, a portfolio is more affected by a few extreme movements in the market than by the sum of many small movements. This motivates risk managers to be primarily concerned with avoiding big unexpected losses. The tool to perform inference  over such unexpected events is \textit{extreme value theory} (EVT) which provides a coherent probabilistic framework to model the tail of a distribution. Standard EVT methods require returns to be independent and identically distributed and their application is based on a number of assumptions which are usually hard to verify in practice. 

 Extreme value mixture models \citep{Scarrott2012} have been introduced to overcome this second deficiency of EVT. These do not require any arbitrary assumption. Although some non-stationary extensions exist \citep[e.g.][]{Nascimento2016}, such models are not capable of explicitly taking into account the structure of financial returns which are often destabilized by shocks concurring with periods of different extreme behaviors.

We introduce here a new class of models, termed \textit{changepoint extreme value mixture models}, which, whilst not requiring any of the arbitrary assumptions usually made in EVT, are also able to formally represent different extreme regimes caused by financial shocks. We demonstrate below that this approach not only correctly identifies the location of such shocks, but also gives  model-based uncertainty measures about these. 
 
Inference is carried out within the \textit{Bayesian} paradigm using the MCMC machinery \citep{Gamerman2006}, enabling us to straightforwardly deliver a wide variety of estimates and predictions of quantities of interest, e.g. high quantiles.

Before formally defining our approach, univariate EVT and non-stationary (extreme) models are reviewed to highlight the relevance and the novelty of our methodology.

\subsection{Extreme value theory}
A common approach to model extremes, often referred to as \textit{peaks over threshold},  studies the exceedances over a threshold. A key result to apply this methodology is due to \citet{Pickands1975} which states that if a random variable X with endpoint $x_e$ is in the  domain of attraction of a generalized extreme value distribution  then $\lim_{u\rightarrow x_e}\mathbb{P}(X\leq x+u|X>u)=G(x)$, where $G$ is the distribution function (df) of the  \textit{generalized Pareto distribution} (GPD). The df $G$ is defined as
\begin{equation*}
G(x|\xi,\sigma,u)=\left\{
\begin{array}{ll}
1-\left(1+\xi\frac{x-u}{\sigma}\right)^{-1/\xi},& \mbox{if } \xi\neq 0,\\
1-\exp\left(-\frac{x-u}{\sigma}\right), & \mbox{if }\xi=0,
\end{array}
\right.
\end{equation*}
for  $u,\xi\in\mathbb{R}$ and $\sigma\in\mathbb{R}_{+}$, where the support is $x\geq u$ if $\xi\geq 0$ and $0\leq x\leq u-\sigma/\xi$ if $\xi<0$. Therefore, the GPD is bounded if $\xi<0$ and unbounded from above if $\xi\geq 0$. The application of this result in practice entails first the selection of a threshold $u$ beyond which the GPD approximation appears to be tenable and then the fit of a GPD over data points that exceed the chosen threshold. Thus  only a small subset of the data points, those beyond the chosen threshold, are formally retained during the inferential process. 

The choice of the threshold over which to fit a GPD is hard and arbitrary. Although  tools to guide this choice exist \citep{Davison1990,Dumouchel1983}, inference can greatly vary for different thresholds \citep{Scarrott2012,Tancredi2006}.

\subsection{Extreme value mixture models}
To overcome the difficulties associated with the selection of  a threshold, a variety of models called \textit{extreme value mixture models} \citep[][]{Scarrott2012} have been recently defined, which formally use the full dataset and do not require a fixed threshold. These  combine a flexible model for the bulk of the data below the threshold, a formally justifiable distribution for the tail and uncertainty measures for the threshold.

\begin{figure}
\begin{center}
\includegraphics[scale=0.3]{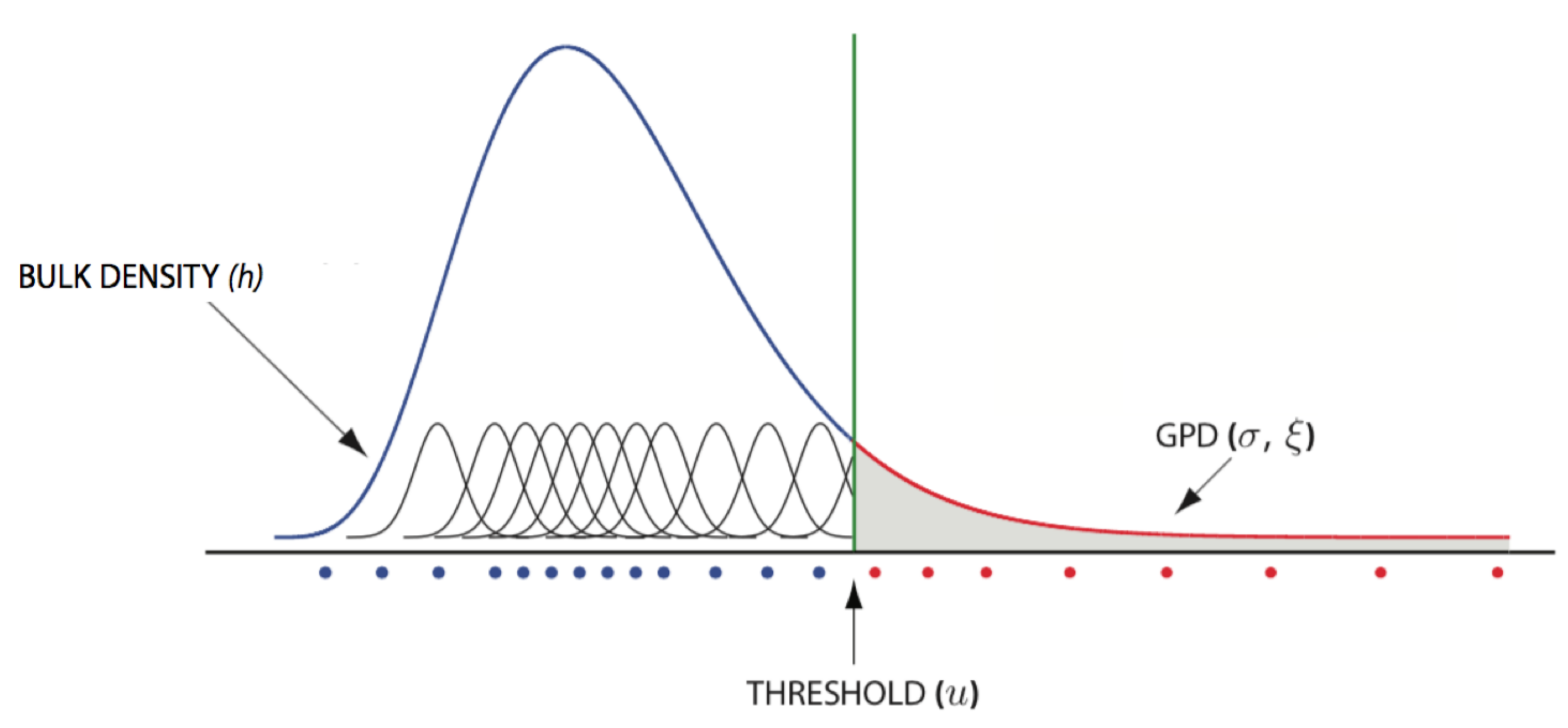}
\end{center}
\vspace{-0.5cm}
\caption{Distribution of an extreme value mixture model with bulk density $h$ and a GPD as tail density \citep[from][]{Scarrott2012}. \label{fig:evmm}}
\end{figure}

The density function $f$ of an extreme value mixture model can be generally defined  as
\begin{equation}
f(x|\Phi,\Psi)=
\begin{cases}
h(x|\Phi), & x \leq u
\\
[1-H(u|\Phi)] g(x|\Psi), & x > u 
\end{cases},
\label{eq:evmm}
\end{equation}
where $h$ is the density, parametrized by $\Phi$, of the bulk, i.e. the portion of data below the threshold $u$,  $H$ its is df and $g$ is the GPD density function with parameters $\Psi=\{\xi,\sigma,u\}$, which models the tail of the distribution above the threshold $u$. Figure \ref{fig:evmm} illustrates the typical form of an extreme value mixture model using a flexible model $h$ for the bulk of the distribution, often defined as a mixture of density functions.

The first proposal to use the full dataset to estimate both the threshold location and the tail of the distribution is due to \citet{Behrens2004}, which used a Gamma for the bulk. Since then a variety of proposal for the bulk have been used, including a Normal distribution \citep{Carreau2009}, an infinite mixture of Uniforms \citep{Tancredi2006}, a mixture of Gammas \citep{Nascimento2012} and a kernel estimator \citep{Macdonald2011}.

\citet{Nascimento2012} demonstrated that nothing is lost in extreme estimation by using the full dataset in cases where the determination of the threshold is easy. Conversely, when uncertainty about the threshold location is high, extreme value mixture models outperform the standard peaks over threshold approach.

\subsection{Non-stationary extremes}
The above methods assume that all observed data come from a same underlying distribution independently. However, in financial, as well as ecological, applications the structure and amplitude of extremes events usually changes through time. For this reason, inference over financial extremes is often carried out using dynamic models. In this direction, \citet{Bollerslev1987} used a GARCH(1,1) model with Student-T innovations to explicitly take into account of the longer tails often encountered in financial datasets. 

Dynamic models based on EVT then started to appear. For instance, \citet{Mcneil2000} proposed a two-stage approach where dependence is first removed using a GARCH model followed by GPD estimation to the assumed independent residual innovations. In a Bayesian setting, \citet{Huerta2007} proposed a hierchical dynamic model based on the generalized extreme value (GEV) distribution, whilst \citet{Zhao2011} defined a GARCH model directly over the GEV parameters. Dynamic extensions of extreme value mixture models have been recently defined \citep{Lima2018r,Nascimento2016}, but in our experience these often require fine tuning of their parameters to work reliably.

Although the above approaches take into account the time dependent nature of rare events, in financial settings extreme variations occur by sudden shocks caused by exogenous agents as described, for instance, by \citet{Caldara2016} and \citet{Dierckx2010}. Financial returns typically show clusters of observations in the tails, a phenomenon often termed volatility clustering. For this reason, inference can be expected to be more accurate by formally taking into account the nature of financial extreme events. 

Changepoint models allow for changes of the model distribution at multiple unknown time points and therefore can be faithfully used to represent and make inference over financial shocks. Some of the first changepoint models using the Bayesian MCMC machinery are due to \citet{Albert1993} and \citet{Carlin1992}, which were extended to multiple changepoints  in \citet{Stephens1994}. Since then the number of changepoint models proposed in the literature has increased dramatically \citep[see e.g][]{Bauwens2017,Giordani2008,Ko2015}. However, changepoint models which explicitly study distributional changes in the structure of the extremes are very limited. 

In the frequentist setting, \citet{Dierckx2010} and \citet{Jaruvskova2008} defined an hypothesis testing routine to investigate the presence of changepoints in GPD and GEV distributions, respectively. In the Bayesian setting, \citet{Nascimento2017} developed MCMC algorithms to identify changepoints in GEV models. Here we propose a highly flexible, new approach for inference over extremes which not only estimates the location of extreme changepoints but also the structure of the extremes within each regime by using the full dataset and without requiring any ad-hoc assumptions.

\subsection{Outline of the paper}
Our approach and inferential routines are next described in Section \ref{sec:changepoint}. Section \ref{sec:simulation} presents a simulation study to both investigate their performance and address the issue of model choice. In Section \ref{sec:data} our methodology is applied to two real-world financial applications: 2-days maxima absolute returns of the NASDAQ stock and negative daily returns of the Royal Bank of Scotland (RBS) stock. We conclude with a discussion.

\section{Changepoint extreme value mixture models}
\label{sec:changepoint}

\begin{figure}
\begin{center}
\includegraphics[scale=0.5]{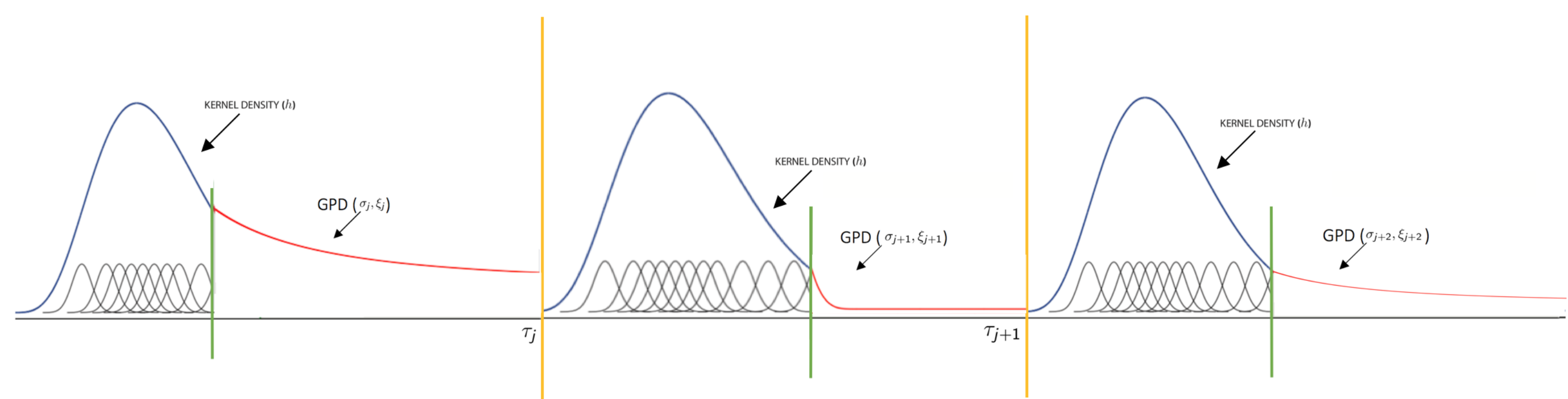}
\end{center}
\vspace{-0.5cm}
\caption{Distribution of a changepoint extreme value mixture model, with common bulk density $h$ and regime dependent GPD tail densities.\label{fig:cevmm}}
\end{figure}

Let $x_1, x_2, \cdots, x_n$ be a series of time-ordered observations. The probability density function of a changepoint extreme value mixture model is defined as
\begin{equation}
f(x_{t}|\Phi,\Psi,\tau)=
\begin{cases}
h(x_{t}|\Phi), & x_t \leq u_j, \quad t \in (\tau_{j-1}, \tau_{j}], \quad j\in [k] \\
[1-H(u_j|\Phi)] g_j(x_{t}|\Psi_j), & x_{t} > u_{j}, \quad t \in (\tau_{j-1}, \tau_{j}], \quad j\in[k] \\
\end{cases}
\label{eq:model}
\end{equation}
where $h$ is a model parametrized by $\Phi$ for the bulk below the threshold $u_{j}$, $H$ its df, $g_j$ a GPD density whose parameters are $\Psi_j = \{u_j,\xi_j,\sigma_j\}$, $\tau=\{\tau_0,\dots,\tau_k\}$ the changepoint locations, $\Psi=\{\Psi_1,\dots,\Psi_k\}$ and $[k]=\{1,\dots,k\}$. The parameters of the GPD vary according to the regime in which the observations above the regime-dependent threshold are situated, whilst the bulk distribution $h$ is common to all regimes and does not vary.  Thus the changepoints mark a distributional change in the extremes only, and not on the overall distribution of the data.  The changepoints are integer values corresponding to the index of an observations that mark a sudden change in the distribution of the data. In this setting, $\tau_0=0$ and $\tau_k=n$: thus there are $k-1$ inner changepoints and $k$ extreme regimes. Figure \ref{fig:cevmm} gives an illustration of the newly defined model class: whilst the bulk distribution is common to all regimes, the GPD distribution for the tail changes between regimes, alternating between periods of heavy and light tails.

Changepoint extreme value mixture models have the very useful property of a parametric closed form for expected return levels above the threshold in each regime. The expected return level for each $t$ period in time is defined as the $1 - \frac{1}{t}$ quantile, i.e. the value $r_{t}$ for which an equal or higher value is expected to occur once every $t$ periods of time.  From \citet{Nascimento2012},  a return $r_{j,t}$ above the threshold in regime $j$ is given by 
\begin{equation} \label{eq:return}
r_{j,t} = u_{j} + \frac{\sigma_{j}}{\xi_{j}}((1-p_{j}^{*})^{-\xi_{j}}-1) \quad \text{where} \quad p_{j}^{*}= \frac{1-\frac{1}{t}-H(u_{j}|\Phi)}{1-H(u_{j}|\Phi)}.
\end{equation}

The model definition in equation (\ref{eq:model}) is general and for practical purposes it needs to be refined by a specific choice of density $h$. Next we present two possible choices based on finite mixtures that we  use in our applications in Section \ref{sec:data}, but in general $h$ can be any density over which Bayesian inference can be carried out.

\subsection{The CMGPD model}
When the common distribution $h$ for the bulk is a finite mixture of Gammas, we say that the changepoint extreme value mixture model is a CMGPD$_l^k$ model, where $l$ denotes the number of mixture components and $k$ the number of different extreme regimes. The CMGPD model extends the MGPD of \citet{Nascimento2012} to include extreme changepoints. A finite mixture of $l$ Gammas is defined as $h(x_t|\Phi)=\sum_{i\in[l]}p_if_{\textnormal{G}}(x_t|\mu_i,\eta_i)$, where $f_\textnormal{G}$ is a Gamma density parametrized by the mean $\mu_i$ and the shape $\eta_i$, i.e.
\[
f_{\textnormal{G}}(x_t|\mu_i,\eta_i)=\frac{(\eta_i/\mu_i)^{\eta_i}}{\Gamma(\eta_i)}x_t^{\eta_i-1}\exp\{-(\eta_i/\mu_i)x\}, \mbox{ for } x_t>0,
\]
with $\mu_i,\eta_i\in\mathbb{R}_{+}$  and $p_i\in [0,1]$ such that $\sum_{i\in[l]}p_i=1$. The parametrization in terms of mean and shape parameters is used to solve identifiability issues \citep{Wiper2001}. In this setting $H(x|\Phi)=\sum_{j\in[l]}p_jF_{\textnormal{G}}(x|\Phi)$, where $F_{\textnormal{G}}$ is the Gamma df. The CMGPD model can be used to fit data over the positive real line, as for instance absolute financial returns.

The bulk density $h$ could be straightforwardly extended to an infinite mixture model \citep[using e.g. the approach of][]{Fuquene2015}, but this is not required: as demonstrated in \citet{Dey1995} and \citet{Rousseau2011} only as small number of mixture components have non-zero weights in practical applications. Furthermore, in our experience, for financial returns one component only is usually necessary. 
 
\subsection{The CMNPD model}
The CMNPD model is similarly defined to the CMGPD, with the difference that the bulk distribution is now a finite mixture of normal distributions. Formally, $h(x_t|\Phi)=\sum_{j\in[l]}p_jf_{\textnormal{N}}(x_t|\mu_j,\delta_j^2)$, where $f_{\textnormal{N}}(x_t|\mu_j,\delta_j^2)$ is the normal density with mean $\mu_j\in\mathbb{R}$ and variance $\delta_j^2\in\mathbb{R}_{+}$. Thus this model is used in financial applications where interest is in one tail only, for instance to predict negative losses. It extends the model of \citet{Carreau2009} to take into account of distributional extreme changepoints.

\subsection{Prior distribution}
The model definition is completed by assigning prior distributions to the parameters.  GPD parameters of different regimes are a priori assumed independent.  In regime $j$, the prior distribution for $(\xi_j,\sigma_j)$ is the non-informative prior of \cite{Castellanos2007} defined as 
$
\pi(\xi_j,\sigma_j) \propto \sigma_j^{-1}(1+\xi_j)^{-1}(1+2\xi_j)^{-1/2}$.

The priors for the different regimes' thresholds are independent Normal distributions as suggested by \citet{Behrens2004}. The prior means $\mu_u$ are placed around the  90$^{th}$ data quantile while the prior variances $\sigma_u^2$ are chosen so that the 95\% prior credibility interval  ranges a priori from the 50$^{th}$ to the 99$^{th}$ data quantiles, in symbols
$\pi(u_j) = f_{\textnormal{N}}(\mu_u, \sigma_u^2)$.

The changepoints are given an non-informative discrete uniform distribution subject to the restriction $\{\tau_0 < \tau_1 < \cdots  < \tau_k\}$, as suggested by \citet{Stephens1994}:
\begin{equation*}
\pi(\tau_1, \dots, \tau_k) = \frac{1}{\tau_2}\mathbbm{1}_{(1 \leq \tau_1 < \tau_2)} \frac{1}{\tau_3-\tau_1} \mathbbm{1}_{(\tau_1 < \tau_2 < \tau_3)} \cdots \frac{1}{n-\tau_{k-2}} \mathbbm{1}_{(\tau_{k-2} < \tau_{k-1} \leq n)}.
\end{equation*}

The prior distribution for the bulk density parameter $\Phi$ depends on the model used. In both cases the weights $(p_1,\dots,p_l)$ are given a Dirichlet prior with parameter $(1,\dots,1)$. For the CMGPD model, the parameters of the Gammas are non-informative and given as in \citet{Nascimento2012}. Each shape parameter $\eta_j$ is given an independent Gamma prior $\pi(\eta_j)=f_G(\eta_j|c_j/d_j,c_j)$, where $c_j,d_j\in\mathbb{R}_{+}$ are chosen to give a large prior variance. The prior for the Gamma means is $\pi(\mu_1,\dots,\mu_l)=K\prod_{j\in[l]}f_{\textnormal{IG}}(\mu_{j}|a_{j},b_{j})\mathbbm{1}_{(0<\mu_{1}<\cdots <\mu_{l})}$,
where $f_{\textnormal{IG}}$ is the inverse Gamma density, 
$K$ is a normalizing constant and $a_j$ and $b_j$ are chosen to give a large prior variance. The order restriction over the means is set to ensure identifiability.

For the CMNPD model,  priors for the normal mixture parameters $(\mu_j,\delta_j^2)_{j\in[l]}$ are given as follows. The prior for the means is given conditionally on the variances as $\pi(\mu_1,\dots,\mu_l|\delta_1,\dots,\delta_l)=\prod_{j\in[l]}f_{\textnormal{N}}(\mu_j|j/M,(\alpha/\delta_j)^2)\mathbbm{1}_{(\mu_1<\cdots<\mu_l)}$, where $\alpha$ is chosen to give a large prior variance and $M=\max(x_1, \cdots, x_n)$.  This choice is motivated by the symmetry of financial returns, so to assure the closeness of the means to 0, and by identifiability issues. Each mixture variance is independently given a Gamma distribution, i.e. $\pi(\delta^2_j)=f_{\textnormal{G}}(\delta^2_j|c_{j}/d_{j},c_{j})$ where again hyperparameters are chosen to be non-informative. 

The overall prior for a changepoint extreme value mixture model can be written as 
$
\pi(\Phi,\Psi,\tau)=\pi(\Phi)\pi(\tau)\prod_{j\in[k]}\pi(\xi_j,\sigma_j)\pi(\mu_j)
$. 
All priors used are non-informative, giving enough flexibility for the influence of the likelihood in the estimation process.

\subsection{Posterior inference}
For a sample $x=(x_1,\dots,x_n)$ the log-posterior of the CMGPD$_l^k$ model is
\begin{equation}
\label{eq:logpost}
\begin{aligned}
\log \pi(\Phi,\Psi,\tau|x) &\propto \sum_{j\in[k]} \sum_{t:x_t\leq u_j}\log \biggl(\sum_{z\in[l]}p_zf_{\textnormal{G}}(x_t|\mu_z,\eta_z)\biggr) \mathbbm{1}_{(t \in (\tau_{j-1}, \tau_{j}])} \\ &+ \sum_{j\in[k]} \sum_{t:x_t>u_j}\log \biggl(1-\sum_{z\in[h]}p_zF_{\textnormal{G}}(u_j|\mu_z,\eta_z)\biggr) \mathbbm{1}_{(t \in (\tau_{j-1}, \tau_{j}])} \\ 
&+ \sum_{j\in[k]}\sum_{t:x_t > u_j} \log(g(x_t|\Psi_j)) \mathbbm{1}_{(t \in (\tau_{j-1}, \tau_{j}])}+\log(\pi(\Phi,\Psi,\tau))
\end{aligned}
\end{equation}

For the CMNPD$_l^k$ model the log-posterior can be easily deduced by substituting $f_{\textnormal{G}}$ and $F_{\textnormal{G}}$ in equation (\ref{eq:logpost}) with $f_{\textnormal{N}}$ and $F_{\textnormal{N}}$ respectively.

Inference cannot be performed analytically and approximating MCMC algorithms are used. Parameters are divided into blocks and updating of the blocks follows Metropolis-Hastings steps since full conditionals have no recognizable form. Proposal variances are tuned via an adaptive algorithm as suggested in \citet{Roberts2009}. Details are given in Appendix \ref{appendix}. All algorithms are implemented in R.

Summaries of financial extreme returns can be straightforwardly computed from the posterior distribution. Common measures used for financial losses are the Value-at-Risk (VaR) and the expected shortfall (ES).  VaR is generally defined as the risk capital sufficient to cover losses from a portfolio over a holding period of a fixed number of days. It corresponds to the  $p^{th}$ quantile  over a certain time horizon and is denoted as VaR$_p$.  ES, or tail conditional expectation, is defined as the potential size of a loss exceeding a specif VaR$_p$. It corresponds to the expectation conditional on observing values larger than VaR$_p$. For changepoint extreme value mixture models, the expected shortfall in the $j$-th regime takes the closed form
\begin{equation}
\label{eq:ES}
ES_p=\frac{VaR_{p,j}^t}{1-\xi_j}+\frac{\sigma_j-\xi_ju_j}{1-\xi_j},
\end{equation}
where $VaR_{p,j}$ is the Value-at-Risk in regime $j$.

Both VaR and ES are highly non-linear functions of the model's parameters (see equations (\ref{eq:return}) and (\ref{eq:ES})). Thus their posterior distribution cannot be derived analytically. However, the MCMC machinery enables us to derive an approximated distribution for \textit{any} function of the models’ parameters, as demonstrated in our applications in Section \ref{sec:data}.

\section{Simulation study}
\label{sec:simulation}
A simulation study is conducted next with two main purposes: first, to assess the identifiability of the models proposed; second, to validate model selection criteria. For brevity we report here the results for data generated from CMGPD and MGPD, but the same results were observed for CMNPD and MNPD. Two samples of 5000 observations were generated, one from a MGPD$_2$, the other from a CMGPD$_2^3$, where the subscript denotes the number of mixture components and the superscript the number of extreme regimes. In both datasets the mixture parameters  are $(\mu_1,\mu_2)=(2,8)$, $(\eta_1,\eta_2)=(4,8)$ and $(p_1,p_2)=(2/3,1/3)$. For the MGPD data,  GPD parameters were fixed at $\xi=0.4$ and $\sigma=2$, whilst the threshold was placed at the $85^{th}$ theoretical quantile of the Gamma mixture (7.99).

The simulated observations from CMGPD$_2^3$ had changepoint locations $\tau=\{2000,3500\}$. The regime-dependent GPDs were chosen so that $(\xi_1,\xi_2,\xi_3)=(-0.4,0,0.4)$, $(\sigma_1,\sigma_2,\sigma_3)=(0.5,1,1.5)$ and the regimes' thresholds were placed respectively at the $80^{th}$ (6.99), $85^{th}$ (7.99) and $90^{th}$ (9.22) theoretical quantiles. 

Simulations were run on a PC with processor 2,7 GHz Intel Core i5 and 8 Gb RAM. For all simulations, the codes ran for 15000 iterations, with a burn-in of 5000 and thinning every 10, giving a posterior sample of 1000. Convergence was assessed by running parallel chains with different starting values and then comparing the resulting estimates. Details about these can be found online\footnote{Posterior samples from the simulation study, as well as from the real data applications reported in Section \ref{sec:data}, are available at the following links:  \url{https://lattanzichiara.shinyapps.io/CMGPDdt2chains/} (CMGPD$_2^2$ data), \url{https://lattanzichiara.shinyapps.io/MGPD2chains/} (MGPD$_2$ data), \url{https://lattanzichiara.shinyapps.io/CMGPD-NDAQ/} (NASDAQ data) and \url{https://lattanzichiara.shinyapps.io/CMNPD-RBS/} (RBS data).}.

\begin{figure}
  \begin{minipage}[b]{0.49\textwidth}
   \centering
\includegraphics[scale=0.2]{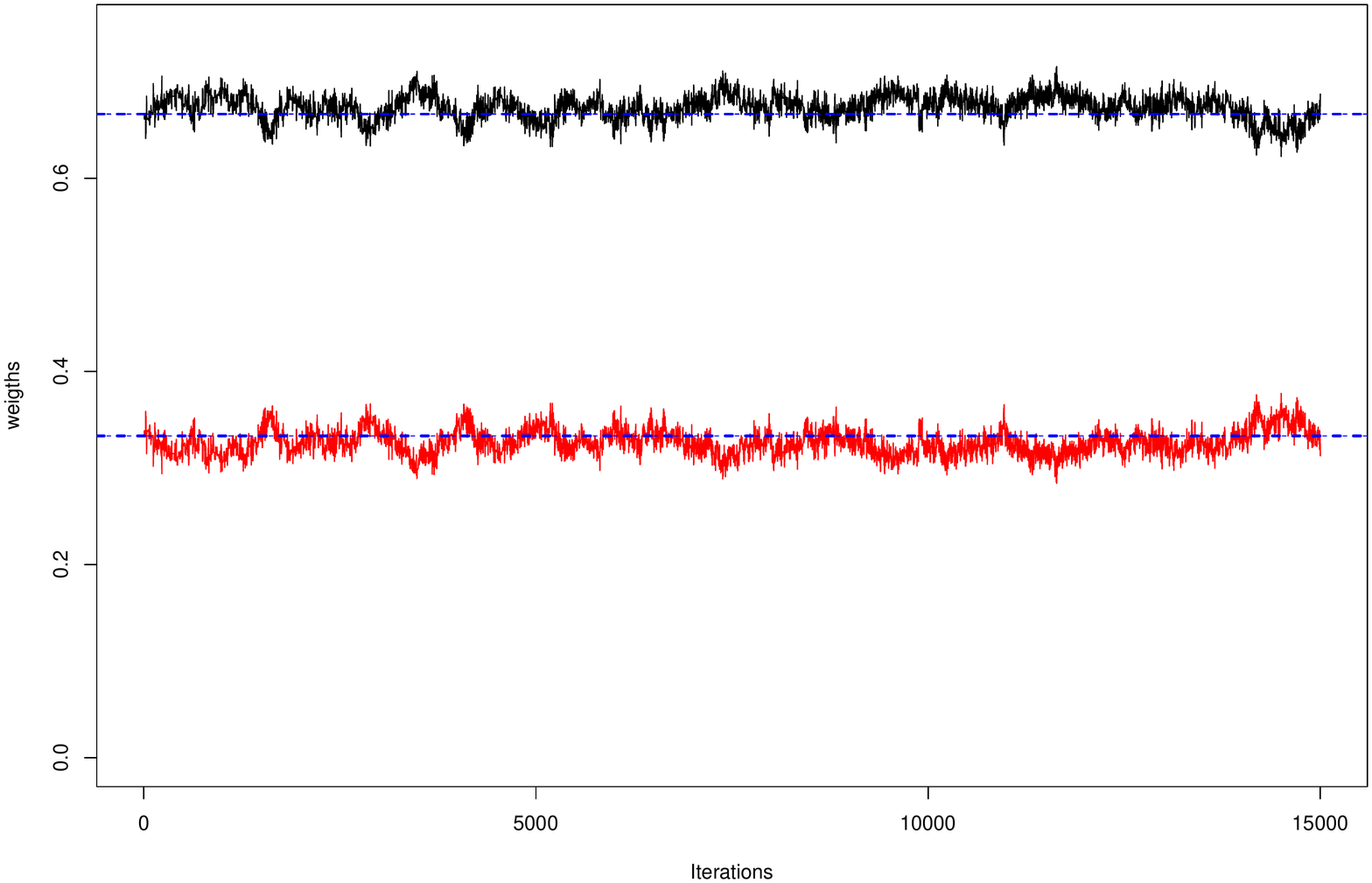}
  \end{minipage}
  \hfill
\begin{minipage}[b]{0.49\textwidth}
\centering
\includegraphics[scale=0.2]{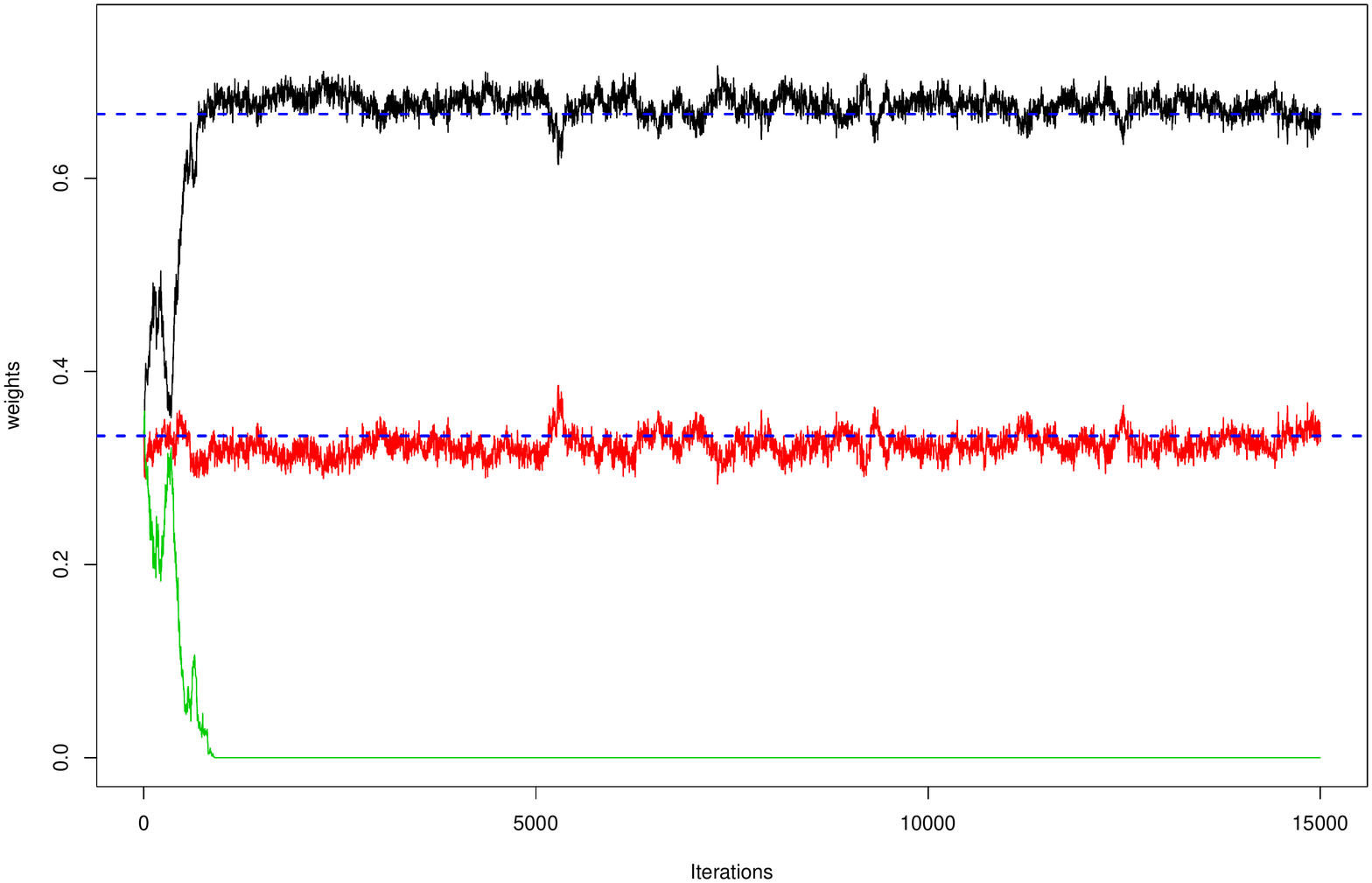}
\end{minipage}
\vspace{-0.3cm}
\caption{Traceplots of the weights $p$ of a CMGPD$_{2}^3$ (left) and a CMGPD$_{3}^3$ (right) fitted to CMGPD$_{2}^3$ simulated data - dashed lines correspond to the true parameter values. \label{fig:traceplots}}
\end{figure}

 In all cases, to reduce the number of models to be compared, the number of mixture components was first chosen by fitting MGPD$_l$ and CMGPD$_l^3$  for various $l$. As already shown in \citet{Nascimento2012} and \citet{Leonelli2017}, the correct number of mixture components can be retrieved from the posterior sample since the weights of all extra components are estimated as zero.  This is demonstrated in Figure \ref{fig:traceplots} where the weight of the third component quickly converges to zero.

\begin{table}
\centering
\scalebox{0.6}{
\begin{tabular}{@{}ccccc@{}}
\toprule
Data            & Model           & BIC      & DIC  & WAIC    \\ \midrule
CMGPD$_{2}^{3}$ & MGPD$_{2}$      & 21212.55 & 21148.29 & 22037.79  \\ \midrule
CMGPD$_{2}^{3}$ & CMGPD$_{2}^{2}$ & 20414.07 & 20361.69 & 20366.07\\ \midrule
CMGPD$_{2}^{3}$ & CMGPD$_{2}^{3}$ & \textbf{20309.73} & 20255.33 & \textbf{20262.07} \\ \midrule
CMGPD$_{2}^{3}$ & CMGPD$_{2}^{4}$ & 20320.46 & \textbf{20255.15} & 20263.74 \\ \midrule \midrule
MGPD$_{2}$ & MGPD$_{2}$      & 22654.83 & 22592.46 & \textbf{22595.90} \\ \midrule
MGPD$_{2}$ & CMGPD$_{2}^{2}$ & \textbf{22649.49} & \textbf{22574.54}  & 22597.84 \\ \bottomrule
\end{tabular}}
\caption{Model selection criteria for models estimated over simulated datasets.}
\label{table:simBIC}
\end{table}

\begin{figure}
  \begin{minipage}[b]{0.32\textwidth}
   \centering
\includegraphics[scale=0.17]{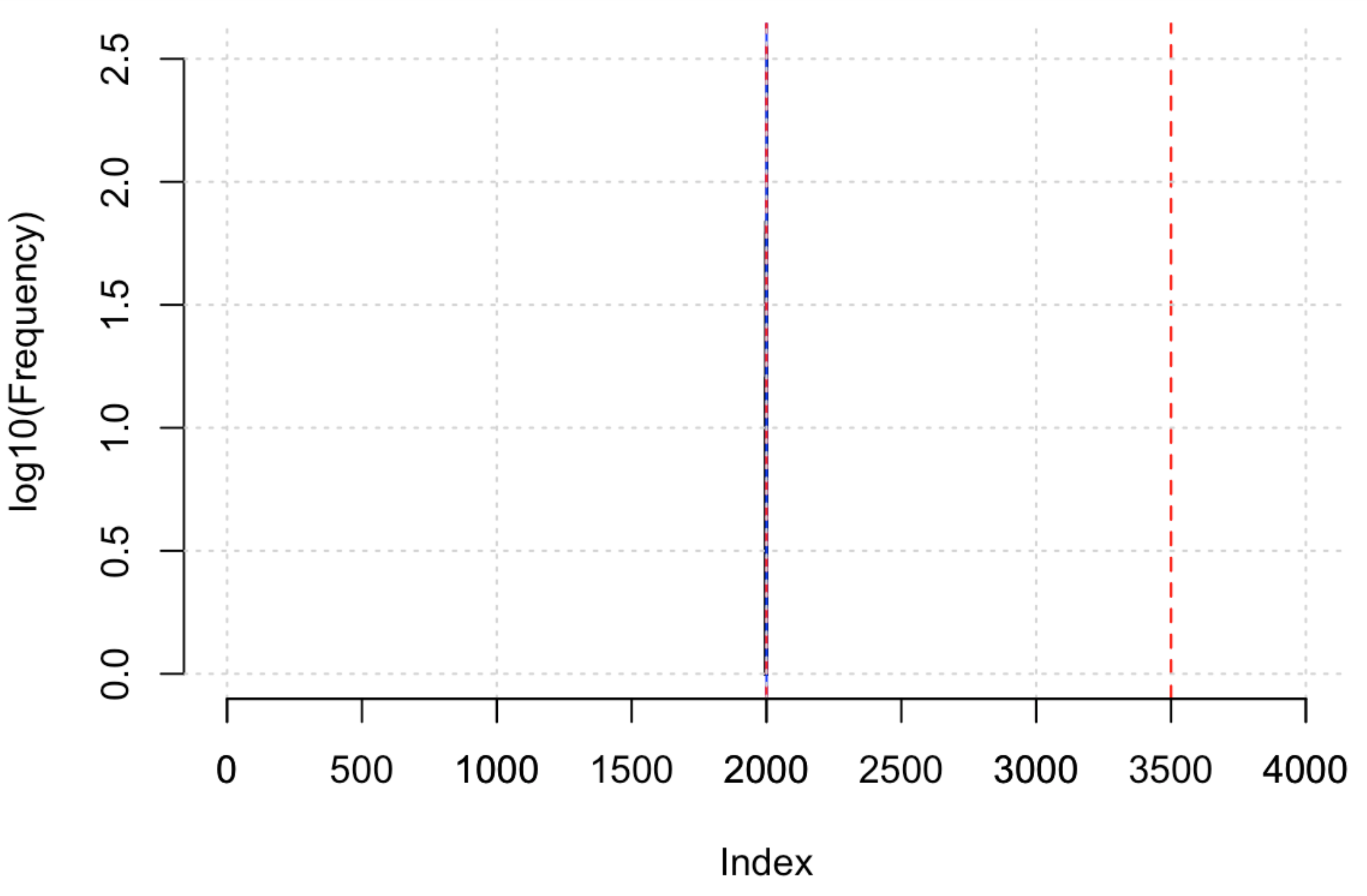}
  \end{minipage}
  \hfill
\begin{minipage}[b]{0.32\textwidth}
\centering
\includegraphics[scale=0.17]{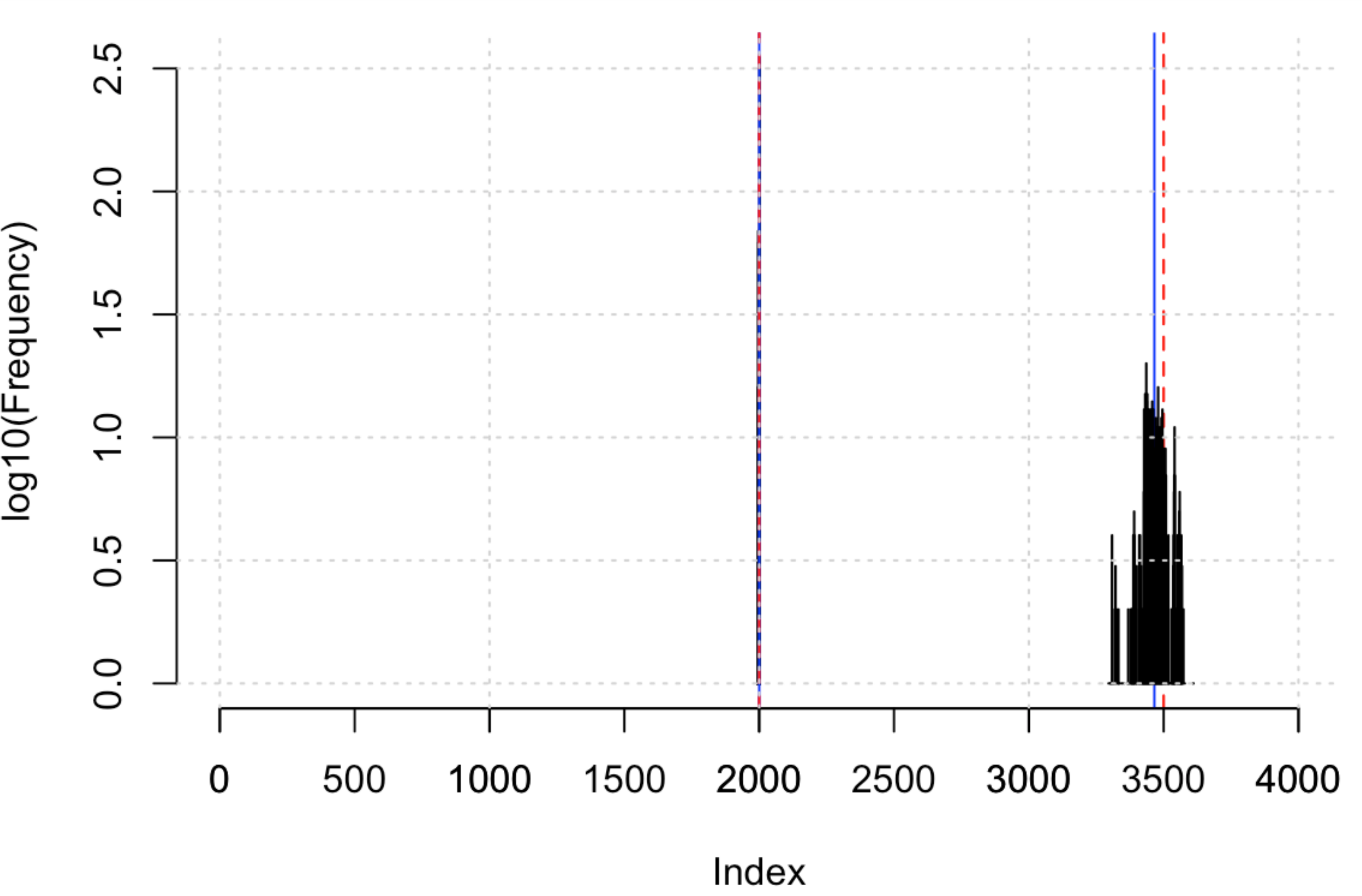}
\end{minipage}
\hfill
\begin{minipage}[b]{0.32\textwidth}
\centering
\includegraphics[scale=0.17]{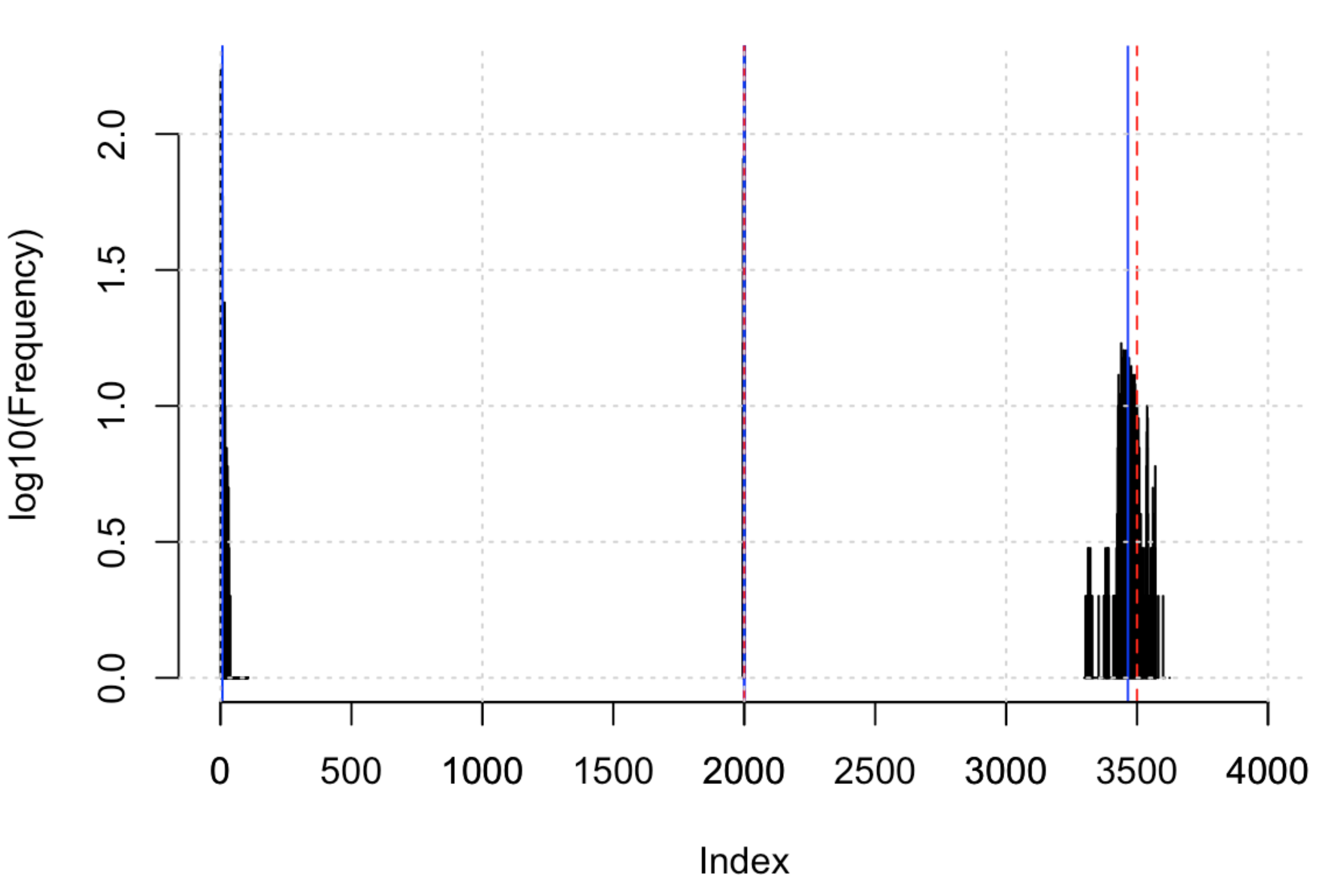}
\end{minipage}
\vspace{-0.3cm}
\caption{Posterior histograms of changepoint locations for CMGPD$_2^2$ (left), CMGPD$_2^3$ (centre) and CMGPD$_2^4$ (right). True values are $\tau_1=2000$ and $\tau_2=3500$, indicated by dashed lines. Full vertical lines: posterior means. \label{fig:changepoint}}
\end{figure}

Having fixed the number of mixtures, models with varying changepoints' numbers were  fitted to the simulated datasets. As already discussed in \citet{Leonelli2017}, standard model selection criteria often fail to identify the correct model in the setting of extreme value mixture models. This is shown in Table \ref{table:simBIC} where BIC \citep{Schwarz1978}  and DIC \citep{Spiegelhalter2002} fail to select the true model. Conversely, the true generating model is always preferred by the WAIC of \citet{Watanabe2010}. This criterion has been shown to be particularly robust for mixtures and non-identifiable models.

\begin{table}[] 
\centering
\scalebox{0.6}{
\begin{tabular}{@{}cccc@{}}
\toprule
Parameter & $\xi_1=-0.4$ & $\xi_2=0$ & $\xi_3=0.4$ \\ \midrule
MGPD$_{2}$ & 0.30 (0.23,0.39)& &    \\ \midrule
CMGPD$_{2}^{3}$ & -0.38 (-0.43,-0.31)  & 0.01 (-0.11,0.18) & 0.49 (0.29,0.76)   \\ \midrule
Parameter & $\sigma_1=0.5$& $\sigma_2=1$ & $\sigma_3=1.5$ \\ \midrule
MGPD$_{2}$ & 1.09 (0.99,1.21) & &    \\ \midrule
CMGPD$_{2}^{3}$ & 0.48 (0.43,0.53) & 0.98  (0.80,1.19) & 1.52 (1.12,1.96)   \\ \midrule
Parameter & $u_1=6.99$ & $u_2=7.99$ & $u_3=9.22$ \\ \midrule
MGPD$_{2}$ & 6.99 (6.99,7.00)  & &     \\ \midrule
CMGPD$_{2}^{3}$ & 6.99 (6.99,7.00)  & 8.02 (8.00,8.04) & 9.07 (8.80,9.25)  \\ \bottomrule
\end{tabular}}
\caption{Posterior means and 95\% credibility intervals for the parameters of simulated CMGPD$_2^3$ data related to the tail densities ($\xi$, $\sigma$ and $u$). \label{table:tail}}
\label{cp2}
\end{table}

The number of regimes can be further identified when fitting CMGPD with non-necessary changepoints since the exceeding locations converge to values very close to $0$, $n$ or another changepoint, depending on the starting values of the MCMC algorithm, as noted in \citet{Nascimento2017}. This can be seen in Figure \ref{fig:changepoint}. When a CMGPD$_2^4$ model is estimated over CMGPD$_2^4$ data, the two changepoints are correctly identified, whilst the third is located close to zero. Conversely, fitting a CMGPD$_2^2$ model over CMGPD$_2^3$ data the only changepoint is estimated around the true changepoint giving a larger distributional change: in this case the one at $t=2000$  associated to a switch from an upper bounded distribution to an unbounded one. The histograms further show that in all cases, uncertainty about the strongest changepoint is limited, whilst the posterior distribution for the changepoint located at $t=3500$ has a larger variance. The same conclusion can be drawn when fitting a CMGPD$_2^2$ model over MGPD$_2$ data, since the posterior mean of the only changepoint is $40.25$ with 95\% credibility interval $(9,88)$.

Having ensured that the true model can be correctly chosen, the identifiability of the parameters is investigated next. As in \citet{Nascimento2012} all bulk parameters are correctly estimated (see the online apps for further details).  But more interestingly, Table \ref{table:tail} demonstrates that tail parameters are well estimated for all regimes when using data simulated from the CMGPD$_2^3$ model. When an MGPD$_2$ model is fitted over this dataset, each tail parameter is estimated around an average value of those of all regimes. When the CMGPD$_2^2$ model is fitted over MGPD$_2$ data, the parameters associated to the non-empty regime well estimate the true tail parameters, as shown in Table \ref{table:tail2}.

\begin{table}[] 
\centering
\scalebox{0.6}{
\begin{tabular}{@{}ccccccc@{}}
\toprule
& $\xi_1$ & $\xi_2$ & $\sigma_1$ & $\sigma_2$ & $u_1$ & $u_2$ \\  \midrule
MGPD$_{2}$ & 0.43 (0.33,0.55)  & & 2.02 (1.76,2.46) && 8.28 (7.82,9.44)  \\ \midrule
CMGPD$_{2}^{2}$ & 0.09 (-0.50,0.59)& 0.45  (0.35,0.59) & 1.72 (0.15,4.01) & 2.04 (1.77,2.52) & 9.07 (7.37,12.93) & 8.39 (8.00,9.37) \\\bottomrule
\end{tabular}
}
\caption{Posterior means and 95\% credibility intervals for the parameters of simulated MGPD$_{2}$ data with true parameters $\xi = 0.4$, $\sigma = 2$ and $u=8.02$. \label{table:tail2}}
\end{table}

Given the use of non-informative priors, there is a clear indication that the likelihood can correctly identify the true values. In particular, the estimation of the tail parameters is highly successful evidencing the ability of the model to recover varying tail behavior.

\section{Applications}
\label{sec:data}
\subsection{NASDAQ absolute daily returns}
The first financial dataset considered consists of daily returns of NASDAQ stock  values from January 1996 to December 2017. Daily returns are considered in their absolute value and, in order to avoid excess return clustering, maxima of sets of 2 days were considered for a total of 2768 observations.  The aim is the estimation of volatility of the composite index over time comparing the MGPD and the CMGPD approaches. 

\begin{table}
\centering
\scalebox{0.6}{
\begin{tabular}{@{}cccc@{}}
\toprule
 Model           & BIC      & DIC  & WAIC    \\ \midrule
 MGPD$_{1}$      & 7527.67 & 7488.93 & 7329.53  \\ \midrule
 CMGPD$_{1}^{5}$ & 7492.97 & 7408.87  & 7203.87 \\ \midrule
CMGPD$_{1}^{6}$ &  \textbf{6696.19}& 6615.71 & \textbf{6776.51} \\ \midrule
CMGPD$_{1}^{7}$ & 6719.61 & \textbf{6612.89}& 7106.92 \\  \bottomrule
\end{tabular}}
\caption{Model selection criteria for models estimated over  NASDAQ data.}
\label{table:nasdaqBIC}
\end{table}

The number of Gamma components for the bulk was first investigated and it was observed that only one component is needed. To choose the number of changepoints we resort to information criteria and posterior locations. The most reliable WAIC criterion favors a model with 6 regimes as reported in Table \ref{table:nasdaqBIC}, which also includes evidence that a changepoint approach outperforms the static one. This is confirmed in Table \ref{table:nasdaqchangepoint} reporting the posterior distribution of the changepoints: the posterior mean of the first CMGPD$_1^7$ changepoint  equals 4, thus giving an empty regime and confirming the optimality of CMGPD$_1^6$.

\begin{table}
\begin{center}
\scalebox{0.6}{
\begin{tabular}{@{}ccccccc@{}}
\toprule
 & $\tau_1$                 & $\tau_2$                 & $\tau_3$              & $\tau_4$              & $\tau_5$              & $\tau_6$              \\ \midrule
CMGPD$_1^5$ & 918 (702,1036)& 1336 (921,1599) & 1602 (1581,1636) & 1645 (1626,1673) & & \\ \midrule
CMGPD$_{1}^{6}$ & 323 (318,327) & 915 (907,929) &1594  (1588,1598) &1681 (1668,1695) &2049 (2008,2094) &   \\ \midrule
CMGPD$_{1}^{7}$ & 4 (1,8) & 323  (318,326)& 915 (907,926)  & 1595  (1590,1598) & 1679 (1667,1690)& 2029  (2006,2092)\\ \bottomrule
 \end{tabular}}
\end{center}
\caption{Posterior distribution of changepoint locations for models estimated over NASDAQ data. \label{table:nasdaqchangepoint}}
\end{table}

\begin{figure}
\begin{center}
\includegraphics[scale=0.35]{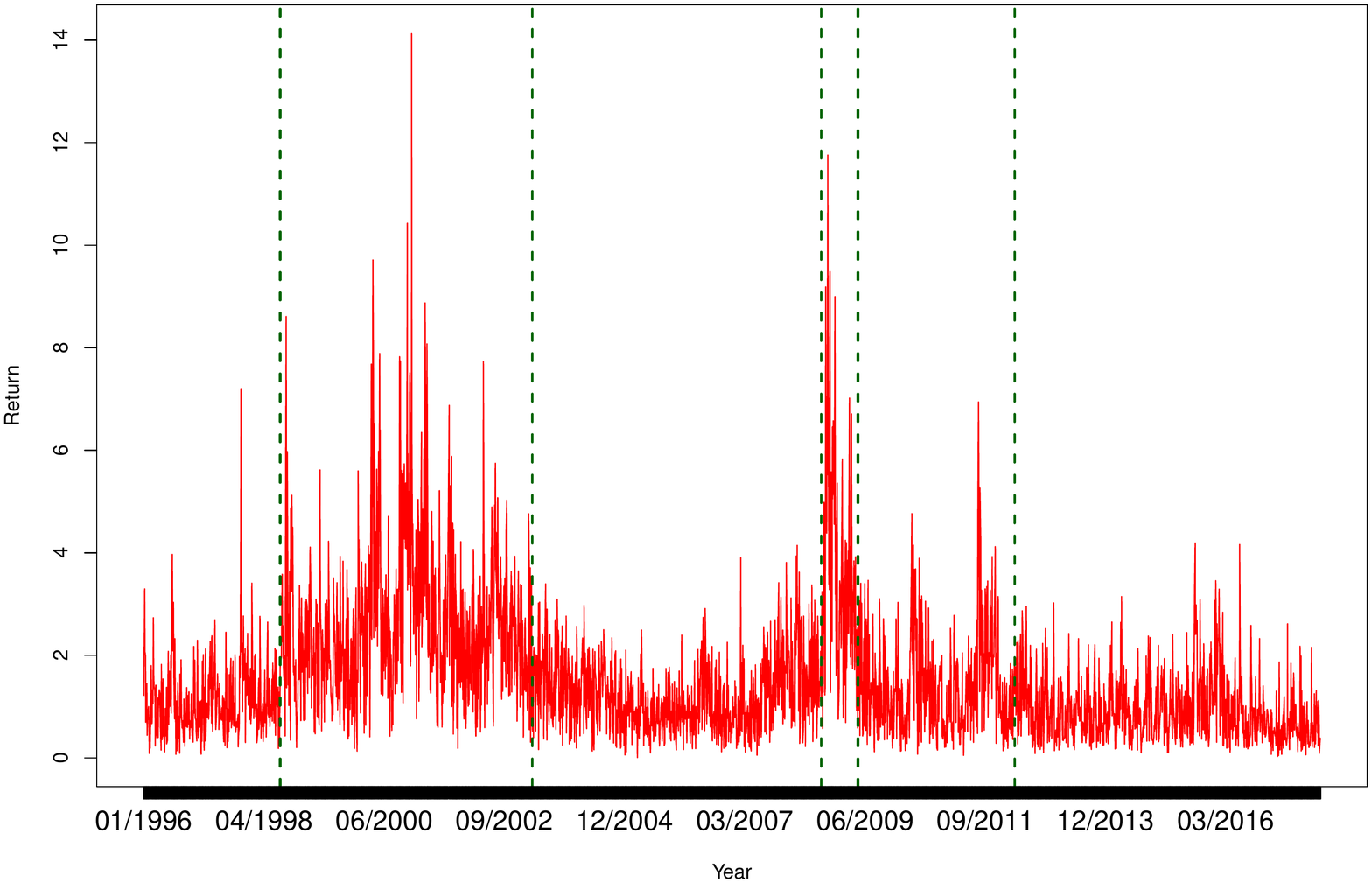}
\end{center}
\vspace{-0.8cm}
\caption{NASDAQ 2-day max absolute returns time series with estimated changepoints using CMGPD$_1^6$. \label{fig:NASDAQ}}
\end{figure}

The posterior means of changepoints  from the CMGPD$_1^6$ model are located on July 1998, April 2003, August 2008, May 2009 and April 2012 as shown in Figure \ref{fig:NASDAQ}. An alternation of regimes with low/medium volatility and high volatility can be noticed, so different tail parameters and returns are to be expected. The regimes with the higher volatility are concurrent to the main events which shook the US stock market in the past 20 years: the second regime show the result of the Dotcom-bubble-burst and the instability after 9/11 while the fourth regime is the direct consequence of the 2008 subprime mortgage crisis. 

Table \ref{table:parNASDAQ} summarizes the posterior distribution of the CMGPD$_1^6$ tail parameters. This demonstrates the flexibility of our approach of discriminating between periods of high and low volatility: in the $2^{nd}$ and $4^{th}$ regimes the estimates of the scale $\sigma$ and shape $\xi$ parameters are larger than all other regimes demonstrating higher level of stress of the market.  The values of the estimated thresholds suggest a particular behaviour of this dataset: the $1^{st}$, $3^{rd}$, $5^{th}$, $6^{th}$ regimes resemble more a GPD distribution than a MGPD. As a result, the estimated thresholds for these regimes are very close to 0. The flexibility of the model proposed enable us to take this into account without any complication. 

\begin{table}
\begin{center}
\scalebox{0.6}{
\begin{tabular}{@{}cccccc@{}}
\toprule
 $\xi_1$ & $\xi_2$& $\xi_3$& $\xi_4$  & $\xi_5$ & $\xi_6$  \\ \midrule
 -0.08 (-0.12,-0.08) & 0.04  (-0.06,0.19)  & -0.30 (-0.35,-0.22)& 0.04 (-0.3,0.63) & -0.12    (-0.20,-0.002)   & -0.14 (-0.20,-0.06) \\ \midrule
  $\sigma_1$ & $\sigma_2$ & $\sigma_3$& $\sigma_4$           & $\sigma_5$ & $\sigma_6$ \\ \midrule
 0.95  (0.83,1.08)  & 1.42  (1.12,1.66)  & 1.25 (1.09,1.42)   & 2.00  (1.04,3.28)                & 1.32     (1.06,1.59) & 0.86(0.78,0.97)       \\ \midrule 
 $u_1$ & $u_2$ & $u_3$ & $u_4$ & $u_5$ & $u_6$ \\ \midrule
 0.30 (0.27,0.39)& 2.00 (1.82,2.44)& 0.32 (0.22,0.43)& 3.31 (2.28,4.05)& 0.44 (0.24,0.65) & 0.22 (0.20,0.23)  \\ \bottomrule
\end{tabular}}
\end{center}
\caption{Posterior distributions of $\xi$, $\sigma$, $u$ for CMGPD$_{1}^{6}$ estimated over NASDAQ data.\label{table:parNASDAQ}}
\end{table}

\begin{figure}[htbp]
  \begin{minipage}[b]{0.49\textwidth}
  \begin{center}
\includegraphics[scale=0.26]{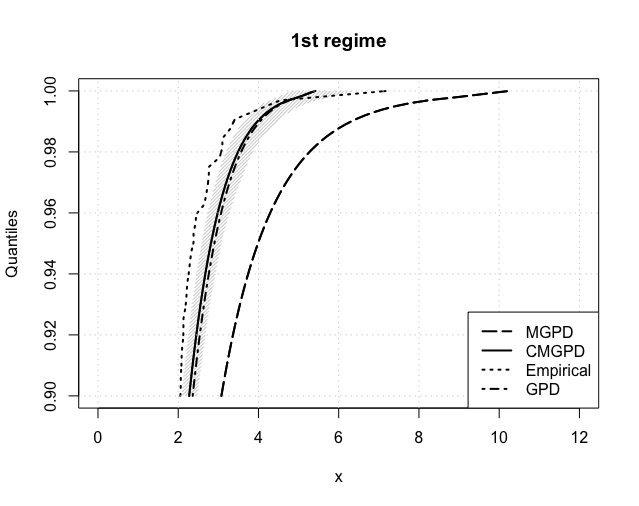}
\end{center}
  \end{minipage}
\hfill
  \begin{minipage}[b]{0.49\textwidth}
    \begin{center}
\includegraphics[scale=0.27]{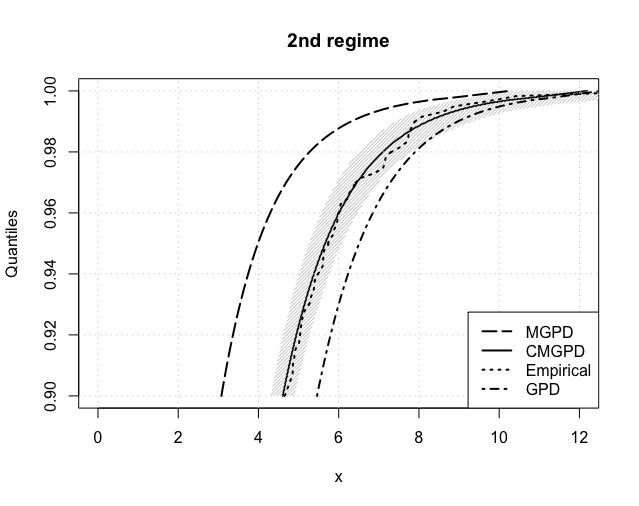}
\end{center}
  \end{minipage}
    \begin{minipage}[b]{0.49\textwidth}
      \begin{center}
\includegraphics[scale=0.27]{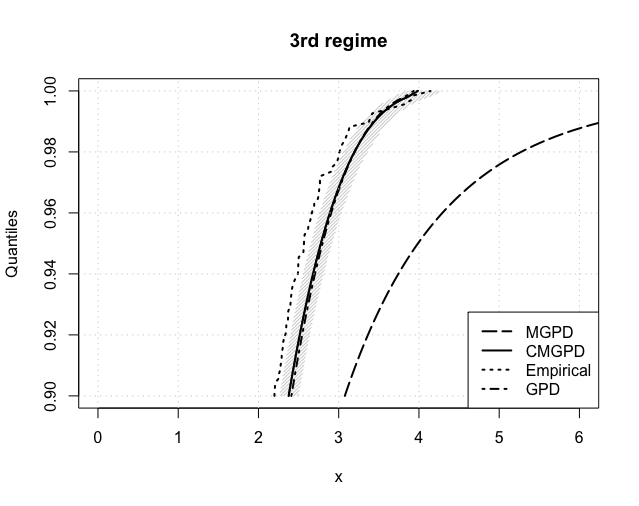}
\end{center}
  \end{minipage}
\hfill
  \begin{minipage}[b]{0.49\textwidth}
    \begin{center}
\includegraphics[scale=0.27]{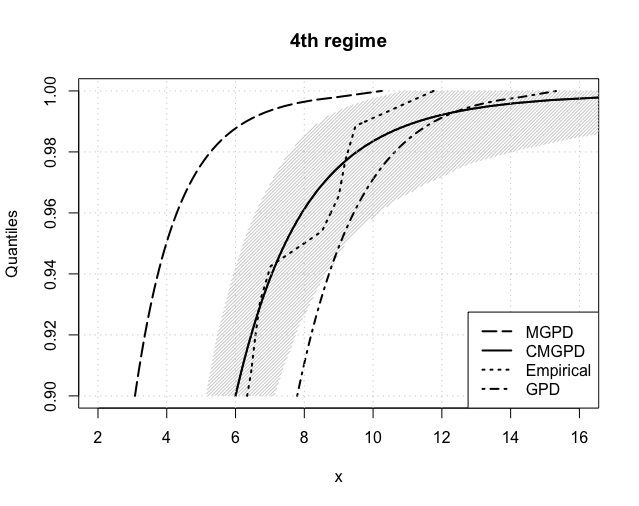}
\end{center}
  \end{minipage}
    \begin{minipage}[b]{0.49\textwidth}
      \begin{center}
\includegraphics[scale=0.27]{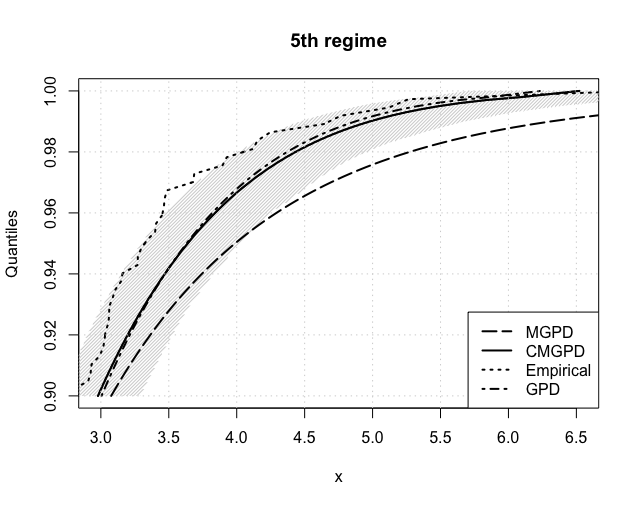}
\end{center}
  \end{minipage}
\hfill
  \begin{minipage}[b]{0.49\textwidth}
  \begin{center}
\includegraphics[scale=0.27]{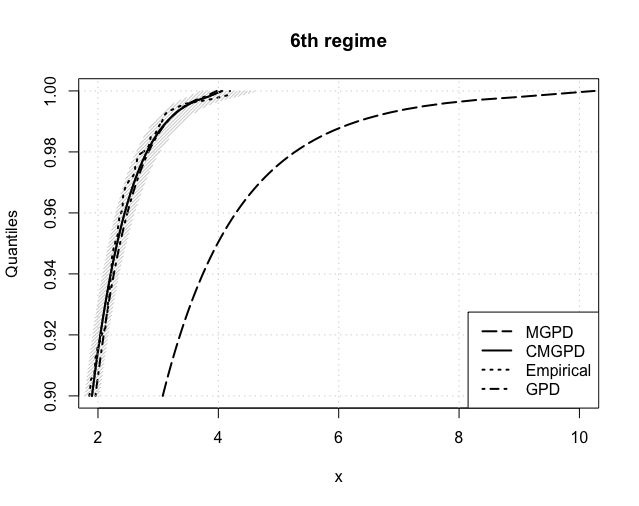}
\end{center}
  \end{minipage}
\caption{Return level plots of each estimated regime for NASDAQ data for a sequence of $t\in [10,1000]$, corresponding to quantiles from 0.90 to 0.999. \label{fig:returnNASDAQ} }
\end{figure}

The expected return levels, for  $t\in [10, 1000]$, are reported in Figure \ref{fig:returnNASDAQ} for each estimated regime. The CMGPD estimates and their 95\% confidence interval (shaded area) are fairly close to the empirical ones in all regimes, thus confirming the goodness of our estimation routines. Returns were further estimated using the MGPD and the GPD (using the threshold estimated by the CMGPD).  The CMGPD estimates are always closer to the empirical values than the MGPD ones. Furthermore the GPD estimates overlays the CMGPD ones in all the regimes with a low threshold, whilst in the others CMGPD clearly outperforms GPD. Thus the use of the full dataset, divided into extreme regimes, leads to better posterior estimates.

\subsection{Royal Bank of Scotland daily returns}

\begin{table}
\centering
\scalebox{0.6}{
\begin{tabular}{@{}cccc@{}}
\toprule
 Model           & BIC      & DIC  & WAIC    \\ \midrule
 MNPD$_{1}$      & 22425.55 & 22422.27 & 22289.93   \\ \midrule
 CMNPD$_{1}^{5}$ & 21510.00  & 21413.04   & 21462.25  \\ \midrule
CMNPD$_{1}^{6}$ &  \textbf{21424.62} & 21356.38 & \textbf{21381.23} \\ \midrule
CMNPD$_{§}^{7}$ & 21429.53 & \textbf{21345.34}& 21392.57 \\  \bottomrule
\end{tabular}}
\caption{Model selection criteria for models estimated over  RBS data.}
\label{table:RBSBIC}
\end{table}

The second financial dataset considered is the Royal Bank of Scotland (RBS) stock daily returns from January 2000 to February 2018 for a total of 4635 observations. In this case positive and negative returns are modeled at the same time and we focus on the estimation of the lower tail (a change of sign is applied for convenience). The estimation efficiency of MNPD and CMNPD models are now investigated. For all models it was first observed that only one Normal component is needed. 

\begin{figure}
\begin{center}
\includegraphics[scale=0.4]{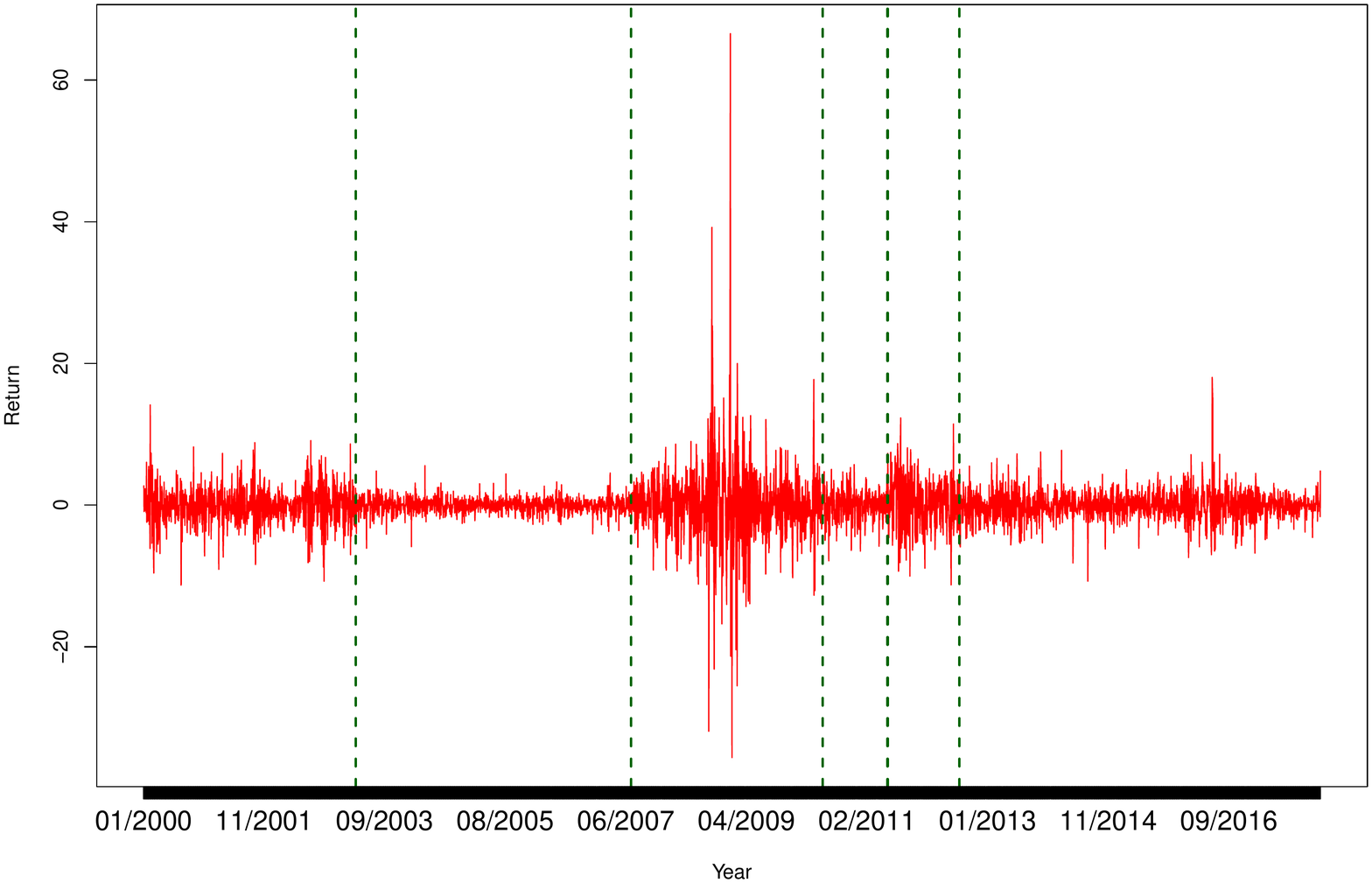}
\end{center}
\vspace{-0.8cm}
\caption{RBS daily negative returns time series with estimated changepoints using CMNPD$_1^6$. \label{fig:RBS}}
\end{figure}

Again all model selection criteria favor a changepoint approach compared to the static one, as reported in Table \ref{table:RBSBIC} and the WAIC  chooses a model with 6 regimes. This is also confirmed by the posterior distribution of the first CMNPD$_1^7$ changepoint,  located at the beginning of the series with a posterior mean of  $61$ and 95\% credibility interval $(36,78)$.

\begin{figure}
\begin{center}
\includegraphics[scale=0.45]{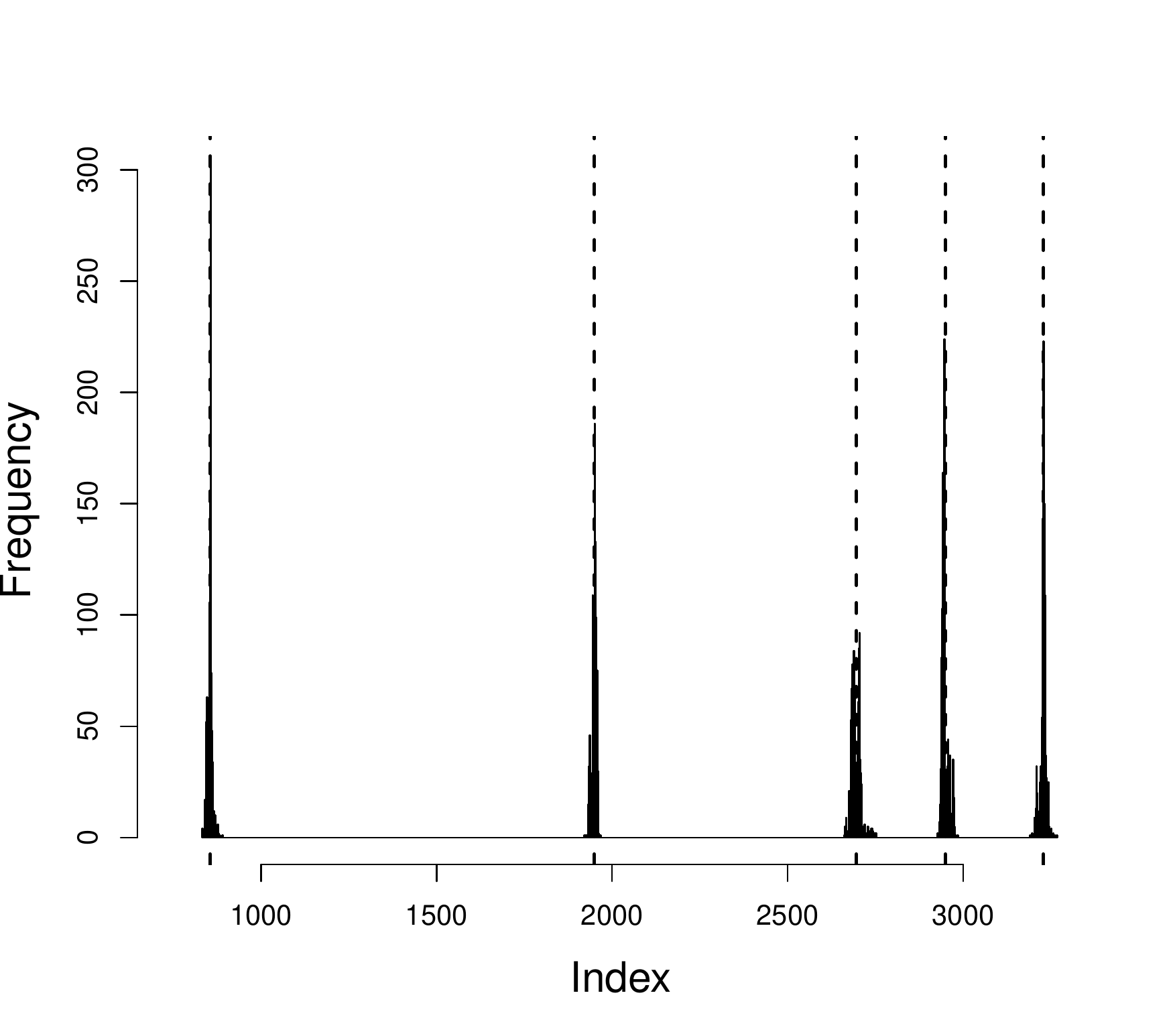}
\end{center}
\vspace{-0.8cm}
\caption{Posterior histograms of changepoint locations for CMNPD$_1^6$ estimated over RBS data. Dashed line denote the posterior means. \label{fig:hist}}
\end{figure}

 The regimes estimated from the CMNPD$_1^6$ model are reported in Figure \ref{fig:RBS}. The  estimated  changepoint are located on April 2003, July 2007, June 2010, June 2011 and August 2012 with posterior distributions shown in Figure \ref{fig:hist}. The regimes show different magnitude in losses, with tail parameters reported in Table \ref{table:RBSpar}. The first and last three regimes represent periods of medium-sized losses, whilts the second one represents a period of high stability. The third regime is by far the most interesting: it is concurrent to the UK’s biggest bank failure in history culminated to the Blue Monday Crash in January 2009. The bank eventually managed to survive thanks to the UK bank rescue package issued by the Government. This is the only regime with a clear heavy tail behaviour ($\xi>0$). The value of $\sigma$ is constant among the regimes, with the exception of the second regime whose value of $\sigma$ indicates lower volatility.

\begin{table}
\centering
\scalebox{0.6}{
\begin{tabular}{@{}cccccc@{}}
\toprule
$\xi_1$ & $\xi_2$ & $\xi_3$ & $\xi_4$& $\xi_5$& $\xi_6$ \\ \midrule
 0.01 (-0.08,0.10) & 0.04 (-0.04,0.13) & 0.37 (0.16,0.61) & -0.12 (-0.31,0.08)&-0.01  (-0.25,0.32)&-0.02 (-0.06,0.03)\\  \midrule
 $\sigma_1$ & $\sigma_2$ & $\sigma_3$ & $\sigma_4$ & $\sigma_5$& $\sigma_6$\\ \midrule
 1.79  (1.57,2.03) & 0.67  (0.58,0.76) & 2.36  (1.79,3.10) & 1.63 (1.23,2.09) &2.07 (1.37,3.09) &1.65 (1.52,1.78) \\ \midrule
 $u_1$ & $u_2$ & $u_3$ & $u_4$& $u_5$& $u_6$ \\ \midrule
-0.008  (-0.02,0.0002) & 0.000  (-0.005,0.0006) & 3.12  (2.23,3.82) & -0.14 (-0.15,0.32) &3.00 (2.31,3.6) &-0.46 (-0.48,-0.43) \\\bottomrule
\end{tabular}
}
\caption{Posterior distributions of $\xi$, $\sigma$, $u$ for CMNPD$_{1}^{6}$ estimated over RBS data.\label{table:RBSpar}}
\end{table}

\begin{figure}
  \centering
  \begin{minipage}[b]{0.49\textwidth}
  \begin{center}
\includegraphics[scale=0.3]{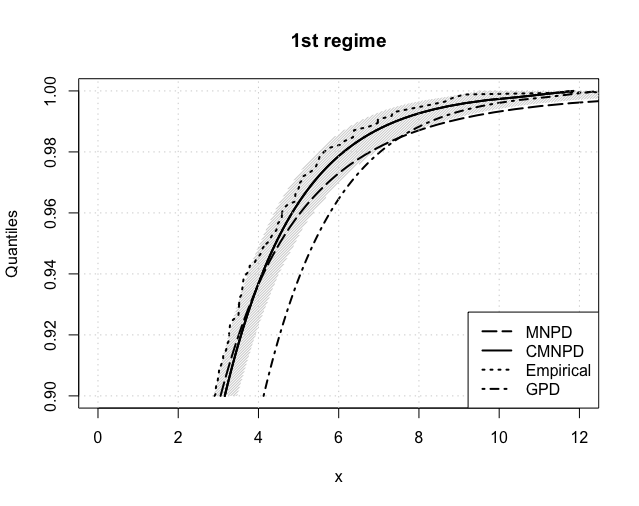}
\end{center}
  \end{minipage}
\hfill
  \begin{minipage}[b]{0.49\textwidth}
  \begin{center}
\includegraphics[scale=0.3]{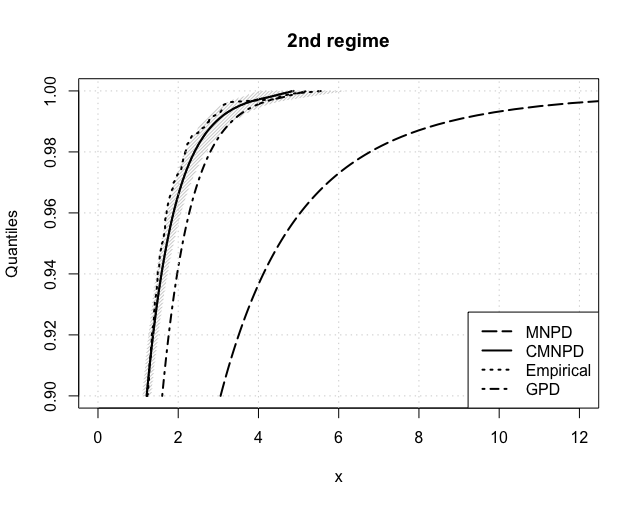}
\end{center}
  \end{minipage}
    \begin{minipage}[b]{0.49\textwidth}
    \begin{center}
\includegraphics[scale=0.3]{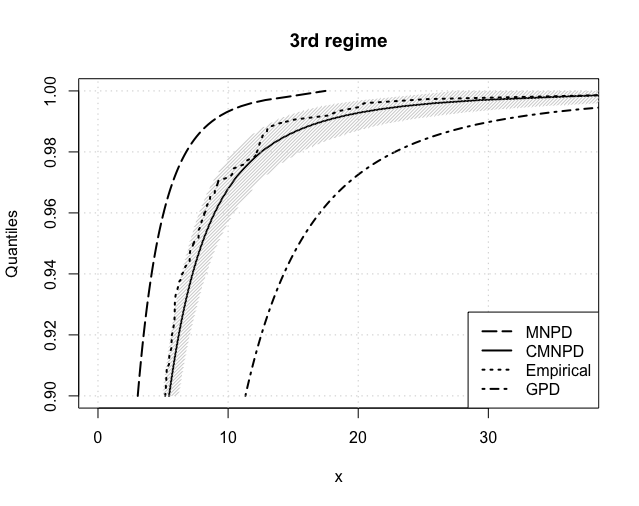}
\end{center}
  \end{minipage}
\hfill
  \begin{minipage}[b]{0.49\textwidth}
  \begin{center}
\includegraphics[scale=0.3]{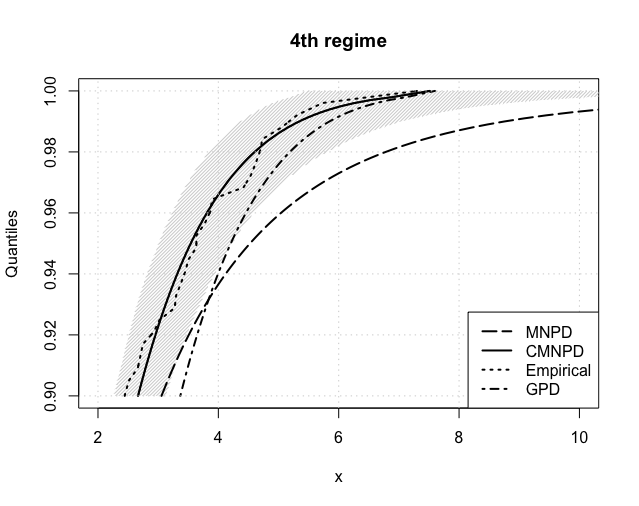}
\end{center}
  \end{minipage}
      \begin{minipage}[b]{0.49\textwidth}
      \begin{center}
\includegraphics[scale=0.3]{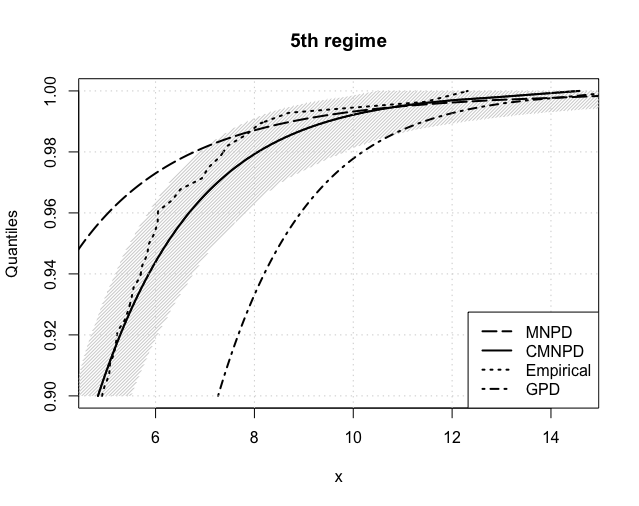}
\end{center}
  \end{minipage}
\hfill
  \begin{minipage}[b]{0.49\textwidth}
  \begin{center}
\includegraphics[scale=0.3]{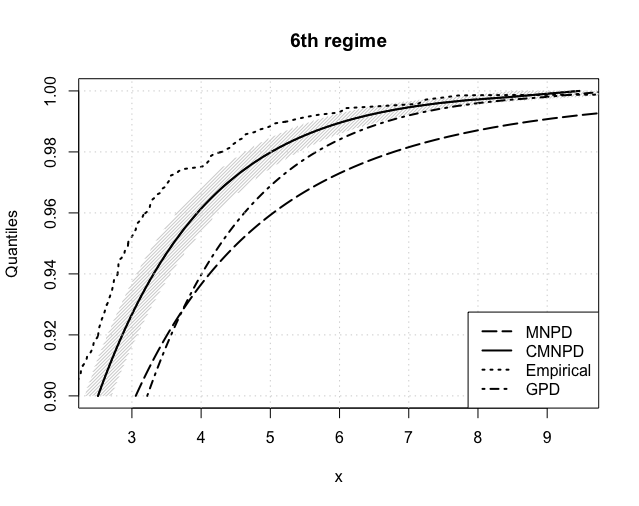}
\end{center}
  \end{minipage}
\caption{RBS VaR computed from 5\% to 0.1\% referred to 1-day time horizon, in each regime. \label{figure:RBSreturn}}
\end{figure}

Figure \ref{figure:RBSreturn} reports the crucial VaR estimates from 5\% to 0.1\% for each regime. The same conclusions can be drawn as in the NASDAQ case, with the CMNPD outperfoming both the MNPD and GPD approaches. Table \ref{table:ES} further summarizes our estimates of the expected shortfall at both 5\% and 1\%. These are compared with the so-called NormFit approach: as reported in \citet{Chang2011} and \citet{Gilli2006} the Basel accord formalizes that financial firms estimate VaR using a normal hypothesis, which is then multiplied by a \lq{s}afety factor\rq{} of 3 to take account of tail's heaviness. Whilst NormFit estimates are comparable to ours at the 5\% level, they highly underestimate risk at the 1\% level.

 The CMNPD model is compared to the GARCH-EVT approach of \citet{Mcneil2000} via \textit{backtesting}: we compare the actual losses at time $t+1$ with the estimated VaR at time $t$. The backtesting is based on a moving window such that at each time $t$, a new set of GARCH(1,1) parameters, residuals and GPD based quantile are estimated. Table \ref{backtab} reports the number of expected VaR$_p$ violations in each regime, equal to $n(1-p)$ with $n$ the number of observations in a regime, and the violations observed from CMNPD and  GARCH-EVT. The CMNPD model always outperforms GARCH-EVT in estimating violations for high-volatility regimes (e.g, the $2^{nd}$ and $3^{rd}$) and in very-high quantiles scenarios (VaR$_{0.5\%}$ and VaR$_{0.1\%}$).  Thus CMNPD better estimates the occurrence of very rare events than the GARCH-EVT approach.

\begin{table}
\begin{center}
\scalebox{0.6}{
\begin{tabular}{@{}cccccccc@{}}
\toprule
Shortfall & Approach  & $1^{st}$ regime & $2^{nd}$ regime & $3^{rd} $ regime & $4^{th}$ regime & $5^{th}$ regime & $6^{th}$ regime \\ \midrule
\multirow{3}{*}{ES$_{5\%}$} & Empirical & 5.87 & 2.30 & 13.15 & 4.63 & 7.40 & 4.50 \\ \cmidrule(l){2-8}
&NormFit & 5.16 & 2.11 & 11.80 & 4.40 & 7.39 & 4.03 \\ \cmidrule(l){2-8}
&CMNPD$_1^6$ & 6.29 (5.56,7.19) & 2.49 (2.23,2.83) & 14.97 (11.50,21.30) & 4.65 (3.85,5.99) & 8.29 (6.98,10.61) & 5.13 (4.73,5.55)\\ \midrule 
\multirow{3}{*}{ES$_{1\%}$} & Empirical & 8.40 & 3.58 & 25.45 & 5.86 & 10.28 & 7.52 \\ \cmidrule(l){2-8}
&NormFit & 6.68 & 2.74 & 15.20 & 5.15 & 9.01 & 5.22 \\ \cmidrule(l){2-8}
&CMNPD$_1^6$ & 9.34 (7.85,11.45) & 3.78 (3.22,4.56) & 30.93 (19.80,55.32) & 6.30 (4.89,9.03) & 10.71 (9.10,18.80) & 7.55 (6.86,8.37)\\ \bottomrule
\end{tabular}}
\end{center}
\caption{ES at 5\% and 1\% estimated using CMNPD$_1^6$ and the NormFit approach. \label{table:ES}}
\end{table}

\begin{table}[ht]
\centering
\scalebox{0.6}{
\begin{tabular}{@{}cccccccc@{}}
\toprule
Value-at-Risk & Approach  & \textbf{$1^{st}$ regime} & \textbf{$2^{nd}$ regime} & \textbf{$3^{rd}$ regime} & \textbf{$4^{th}$ regime} & \textbf{$5^{th}$ regime} & \textbf{$6^{th}$ regime} \\ \midrule
\multirow{3}{*}{VaR$_{5\%}$}&Expected       & 42           & 54            & 37           & 13           & 14           & 70           \\\cmidrule(l){2-8}
&CMNPD          & 40              & \textbf{43}     & \textbf{32}     & \textbf{14}     & 10              & 38              \\ \cmidrule(l){2-8}
&GARCH-EVT      & \textbf{41}     & 40              & 54              & 7               & \textbf{15}     & \textbf{65}     \\ \midrule
\multirow{3}{*}{VaR$_{1\%}$}&Expected       & 8            & 10           & 7            & 2            & 3            & 14          \\\cmidrule(l){2-8}
&CMNPD          & 6               & \textbf{9}      & \textbf{7}      & 2               & 2               & 8               \\\cmidrule(l){2-8}
&GARCH-EVT      & \textbf{9}      & 13              & 16              & 2               & 2               & \textbf{19}     \\ \midrule
\multirow{3}{*}{VaR$_{0.5\%}$}&Expected       & 4            & 5            & 3            & 1          & 1           & 7           \\\cmidrule(l){2-8}
&CMNPD          & \textbf{3}      & \textbf{4}      & \textbf{3}      & \textbf{1}      & \textbf{1}      & \textbf{7}      \\\cmidrule(l){2-8}
&GARCH-EVT      & 8               & 8               & 12              & 2               & 2               & 15              \\ \midrule
\multirow{3}{*}{VaR$_{0.1\%}$}&Expected       & 1            & 1            & 1           & 0           & 0            & 1           \\\cmidrule(l){2-8}
&CMNPD          & \textbf{1}      & \textbf{1}      & \textbf{1}      & \textbf{0}      & \textbf{0}      & \textbf{2}      \\\cmidrule(l){2-8}
&GARCH-EVT      & 4               & 3               & 6               & 1               & 1               & 6               \\ \bottomrule
\end{tabular}}
\caption{Comparing VaR violations with expected violations using CMNPD$_1^6$ and GARCH-EVT.}
\label{backtab}
\end{table}

\section{Conclusions}
The financial literature asserts that not only extreme behaviour may change considerably in time, but also that these variations occur by sudden shocks which deeply affect volatility scenarios. This work puts forward a changepoint generalisation of extreme value mixture models with the ability of detecting multiple changepoints in the tail distribution. The inclusion of regime-dependent GPD parameters enables the switch between light and heavy-tailed behaviour explaining well periods of financial stress and market instability.

Due to the semiparametric nature of the models proposed, Bayesian methods are used.  Despite the use of vague prior information, our inferential routines recover the correct parameter values,  whilst giving uncertainty measures about crucial parameters such as thresholds and changepoint locations. Model choice is easily performed due to the inherent ability of the model to detect the number of mixture components for the bulk and, most importantly, the number of changepoints. 

Our approach outperforms in financial applications all the static and dynamic methods considered for comparison. Since financial markets are heavily affected by unexpected and abrupt variations, extreme regimes are well-captured using changepoint tools, identifying periods of changing  volatility.  Return levels, VaR and ES measures are well estimated by our approach, making it a very powerful tool in a real-data context. Their effectiveness in other fields, for instance environmental and medical applications, is yet to be explored. 

Although the number of changepoints is correctly identified by model selection criteria,  models with a different number of changepoints need to be fitted. We are currently exploring approaches to estimate $k$, the number of changepoints, within our MCMC routines. Recent proposals use the hidden changepoint representation of \citet{Chib1998} coupled with a Dirichlet process  \citep[e.g.][]{Ko2015}. However these fail in our context because the acceptance of a new changepoint location is based on a subset of the observations. Because our changepoints discriminate only tail behaviour, such subsets do not include enough information to identify their location. More promising is the development of reversible jump MCMC algorithms \citep{Green1995}, which have already been successfully applied in changepoint applications, although not in the context of extremes.

\bibliographystyle{Chicago}

\bibliography{Bib}

\appendix
\section{MCMC Algorithms} 
\label{appendix}
\subsection{CMGPD} 

Sampling is carried out in blocks with Metropolis-Hastings proposals. A parameter with a superscript $(s)$ denotes its value at the $s$-th iteration of the algorithm. Let $\mu=\{\mu_1,\dots,\mu_l\}, \eta=\{\eta_1,\dots,\eta_l\},$ $p=\{p_1,\dots,p_l\}$, $u=\{u_1,\dots,u_k\}$, $\sigma=\{\sigma_1,\dots,\sigma_k\}$, $\xi=\{\xi_1,\dots,\xi_k\}$ and $\tau=\{\tau_0,\dots,\tau_k\}$. We denote $\xi_{<j}=\{\xi_1,\dots,\xi_{j-1}\}$, $\xi_{\geq j}=\{\xi_j,\dots,\xi_k\}$ and similarly for other parameters. Recall that $\Phi=\{\mu,\eta,p\}$ and $\Psi=\{\xi,\sigma,u\}$. At each iteration $s$, parameters are updated as follows:\\

\textbf{Sampling $\xi$:} The proposal transition kernel for each $\xi_{j}$, $j\in [k]$, where $k$ is the total number of regimes, is given by a truncated Normal $N(\xi_{j}^{(s)}, V_{\xi_{j}})\mathbbm{1}_{\left(-\sigma_{j}^{(s)}(M_{j}^{(s)}-u_{j}^{(s)}), \infty\right)}$
where $V_{\xi_{j}}$ is a variance appropriately chosen to ensure chain mixing and $M_{j}^{(s)}$ is the maximum of the observations in $(\tau^{(s)}_{j-1},\tau_{j}^{(s)}]$. So, $\xi_{j}^{(s+1)} = \xi_j^{*}$ with probability $\alpha_{\xi_{j}}$, where
\begin{equation*}
\alpha_{\xi_j}= \min\left\{1, \frac{\pi(\Theta^{*}|x) f_{\textnormal{N}}(\xi_{j}^{(s)}, V_{\xi_{j}})\mathbbm{1}_{\left(-\sigma_{j}^{(s)}(M_{j}^{(s)}-u_{j}^{(s)}), \infty\right)}   
}{\pi(\tilde{\Theta}|x) f_{\textnormal{N}}(\xi_{j}^{*},V_{\xi_{j}})\mathbbm{1}_{\left(-\sigma_{j}^{(s)}(M_{j}^{(s)}-u_{j}^{(s)}), \infty\right)} }\right\},
\end{equation*}
$\Theta^{*} = \{\Phi^{(s)}, u^{(s)}, \sigma^{(s)}, \xi^{*},\tau^{(s)}\}$, $\xi^{*} = \{ \xi_{<j}^{(s+1)}, \xi_{j}^{*}, \xi_{>j}^{(s)}\}$ and $\tilde{\Theta} = \{\Phi^{(s)},  u^{(s)}, \sigma^{(s)}, \xi_{<j}^{(s+1)},\xi_{\geq j}^{(s)},\tau^{(s)}\}$. 
\\

\textbf{Sampling $\sigma$:} The proposal transition kernel for each $\sigma_{j}$, $j=1\in [k]$, depends on the value of $\xi_{j}^{(s+1)}$. If $\xi_{j}^{(s+1)} \geq 0$, then $\sigma^{*}$ is sampled from the Gamma distribution $G(\sigma_{j}^{(s)}, {\sigma_{j}^{(s)}}^2/V_{\sigma_{j}})$ where $V_{\sigma_{j}}$ is the variance of the proposal distribution appropriately chosen to ensure chain mixing. If $\xi_{j}^{(s+1)} < 0$, then $\sigma_{j}^{*}$ is sampled from a $N(\sigma_{j}^{(s)}, V_{\sigma_{j}})\mathbbm{1}_{(-\xi_{j}^{(s+1)}(M_{j}^{(s)}-u_{j}^{(s)}), \infty)}$. So, $\sigma_{j}^{s+1}=\sigma_{j}^{*}$ with probability $\alpha_{\sigma_{j}}$ where, if $\xi_{j}^{(s+1)} < 0$, 
\begin{equation*}
\alpha_{\sigma_j}= \min\left\{1, \frac{\pi(\Theta^{*}|x) f_{\textnormal{N}}(\sigma_{j}^{(s)},V_{\sigma_{j}})\mathbbm{1}_{(-\xi_{j}^{(s+1)}(M_{j}^{(s)}-u_{j}^{(s)}),\infty)}
}{\pi(\tilde{\Theta}|x) f_{\textnormal{N}}(\sigma_{j}^{*},V_{\sigma_{j}})\mathbbm{1}_{(-\xi_{j}^{(s+1)}(M_{j}^{(s)}-u_{j}^{(s)}),\infty)}}\right\},
\end{equation*}
and if $\xi_{j}^{(s+1)} > 0$,
\begin{equation*}
\alpha_{\sigma_j}=\min\left\{1, \frac{\pi(\Theta^{*}|x) f_{\textnormal{G}}(\sigma_{j}^{(s)}|\sigma_{j}^{*},{\sigma_{j}^{*}}^2/V_{\sigma_{j}})
}{\pi(\tilde{\Theta}|x) f_{\textnormal{G}}(\sigma_{j}^{*}|\sigma_{j}^{(s)},\sigma_{j}^{(s) 2}/V_{\sigma_{j}})
}\right\},
\end{equation*}
$\Theta^{*} = \{\Phi^{(s)},  u^{(s)}, \sigma^{*}, \xi^{(s+1)},\tau^{(s)}\}$, $\sigma^* = \{ \sigma_{<j}^{(s+1)}, \sigma_{j}^{*}, \sigma_{>j}^{(s)}\}$ and $\tilde{\Theta} = \{\Phi^{(s)}, u^{(s)}, \sigma_{< j}^{(s+1)},\sigma_{\geq j}^{(s)}, \xi^{(s+1)},\tau^{(s)}\}$. 
\\

\textbf{Sampling $u$:} The thresholds $u^{*}_{j}$ are sampled from a $N(u_{j}^{(s)},V_{u_{j}}) \mathbbm{1}_{(a_{j}^{(s+1)}, \infty)}$ distribution where $a_{j}^{s+1}$ is the minimum of the observations in $(\tau^{(s)}_{j-1},\tau_{j}^{(s)}]$ if $\xi_{j}^{(s+1)} \geq 0$ and $a_{j}^{s+1} = M_{j}^{(s)} + \sigma_{j}^{(s+1)}/\xi_{j}^{(s+1)}$  if $\xi_{j}^{(s+1)} < 0$. The lower limit of the truncation is chosen to satisfy the sample space of the GPD.  The variance $V_{u_{j}}$ is chosen to ensure appropriate chain mixing. So  $u_{j}^{(s+1)} = u_{j}^{*}$ with probability $\alpha_{u_{j}}$, where
\begin{equation*}
\alpha_{u_{j}} = \min\left\{1, \frac{\pi(\Theta^{*}|x) f_{\textnormal{N}}(u_{j}^{(s)}, V_{u_{j}}) \mathbbm{1}_{(a_{j}^{(s+1)}, \infty)}
}{\pi(\tilde{\Theta}|x) f_{\textnormal{N}}(u_{j}^{*}, V_{u_{j}}) \mathbbm{1}_{(a_{j}^{(s+1)}, \infty)}}\right\},
\end{equation*}
 $\Theta^{*} = \{\Phi^{(s)}, u^{*}, \sigma^{(s+1)}, \xi^{(s+1)},\tau^{(s)}\}$, $u^{*} = \{ u_{<j}^{(s+1)}, u_{j}^{*}, u_{>j}^{(s)}\}$ and $\tilde{\Theta} = \{\Phi^{(s)}, u_{<j}^{(s+1)},u_{\geq j}^{(s)}, \sigma^{(s+1)}, \xi^{(s+1)},\tau^{(s)}\}$. 
\\

\textbf{Sampling $\eta$: } The proposal kernel for $\eta_{z}$, $z\in[l]$, where $l$ is the number of mixture components, is taken as the Gamma distribution $ G(\eta_{z}^{(s)},{\eta_{z}^{(s)}}^2/V_{\eta_{z}})$, where $V_{\eta_{z}}$ is  chosen to ensure appropriate chain mixing. 
So $\eta_{z}^{(s+1)}=\eta_{z}^{*}$ with probability $\alpha_{\eta_{z}}$, where
\begin{equation*}
\alpha_{\eta_{z}} = \min \left\{1, \frac{\pi(\Theta^{*}|x) f_{\textnormal{G}}(\eta_{z}^{(s)}|\eta_{z}^{*},{\eta_{z}^{*}}^2 /V_{\eta_{z}})
}{\pi(\tilde{\Theta}|x) f_{\textnormal{G}}(\eta_{z}^{*}|\eta_{z}^{(s)},{\eta_{z}^{(s)}}^2/V_{\eta_{z}})}\right\},
\end{equation*}
$\Theta^{*} = \{\mu^{(s)}, \eta^{*}, p^{(s)}, \Psi^{(s+1)},\tau^{(s)}\}$, $\eta^{*} = \{ \eta_{<z}^{(s+1)}, \eta_{z}^{*}, \eta_{>z}^{(s)}\}$ and $\tilde{\Theta} = \{\mu^{(s)}, \eta_{< z}^{(s+1)},\eta_{\geq z}^{(s)}, p^{(s)},\Psi^{(s+1)},\tau^{(s)}\}$. 
\\

\textbf{Sampling $\mu$: } The proposal kernel for $\mu_{z}$, $z\in[l]$, is taken as the Gamma distribution $ G(\mu_{z}^{(s)},{\mu_{z}^{(s)}}^ 2/V_{\mu_{z}}) \mathbbm{1}_{(\mu_{1}^{(s+1)} < \cdots < \mu_{z-1}^{(s+1)} < \mu_{z}^{(s)} < \cdots < \mu_{h}^{(s)})}
$
where  $V_{\mu_{z}}$ is chosen to ensure appropriate chain mixing.
So $\mu_{z}^{(s+1)}=\mu_{z}^{*}$ with probability $\alpha_{\mu_{z}}$, where
\begin{equation*}
\alpha_{\mu_{z}} = \min \left\{1, \frac{\pi(\Theta^{*}|x) f_{\textnormal{G}}(\mu_{z}^{(s)}|\mu_{z}^{*},{\mu_{z}^{*}}^2 /V_{\mu_{z}})\mathbbm{1}_{ (\mu_{1}^{(s+1)} < \cdots < \mu_{z}^{*} < \cdots < \mu_{h}^{(s)})}
}{\pi(\tilde{\Theta}|x) f_{\textnormal{G}}(\mu_{z}^{*}|\mu_{z}^{(s)},{\mu_{z}^{(s)}}^ 2/V_{\mu_{z}}) \mathbbm{1}_{ (\mu_{1}^{(s+1)} < \cdots < \mu_{z}^{(s)} < \cdots < \mu_{h}^{(s)})}}\right\},
\end{equation*}
$\Theta^{*} = \{\mu^{*}, \eta^{(s+1)}, p^{(s)}, \Psi^{(s+1)},\tau^{(s)}\}$, $\mu^{*} = \{ \mu_{<z}^{(s+1)}, \mu_{z}^{*}, \mu_{>z}^{(s)}\}$ and $\tilde{\Theta} = \{\mu_{< z}^{(s+1)},\mu_{\geq z}^{(s)}, \eta^{(s+1)}, p^{(s)}, \Psi^{(s+1)},\tau^{(s)}\}$. 
\\

\textbf{Sampling $p$: }The vector of weights is proposed from a Dirichlet $D_{h}(V_{p}p_1^{(s)}, \cdots, V_{p}p_{h}^{(s)})$, where $V_p$ is chosen to ensure chain mixing. So,  $p^{(s+1)}= p^{*}$ with probability $\alpha_p$, where:
\begin{equation*}
\alpha_{p} = \min \left\{1, \frac{\pi(\Theta^{*}|x) f_{\textnormal{D}}(p^{(s)}|p^{*})
}{\pi(\tilde{\Theta}|x) f_{\textnormal{D}}(p^{*}|p^{(s)}) }\right\},
\end{equation*}
$\Theta^{*} = \{\mu^{(s+1)}, \eta^{(s+1)}, p^{*}, \Psi^{(s+1)},\tau^{(s)}\}$ and $\tilde{\Theta} = \{\mu^{(s+1)}, \eta^{(s+1)}, p^{(s)}, \Psi^{(s+1)},\tau^{(s)}\}$.
\\

\textbf{Sampling $\tau$: } The proposal transition kernel for each $\tau_{j}$, $j\in[k-1]$, is given by a truncated Normal 
$ N(\tau_{j}^{(s)}, V_{\tau_{j}})\mathbbm{1}_{(\tau_{j-1}^{(s+1)}, \tau_{j+1}^{(s)})}$,
where $V_{\tau_{j}}$ is a chosen to ensure chain mixing. So, $\tau_{j}^{(s+1)} = \tau_j^{*}$ with probability $\alpha_{\tau_{j}}$, where
\begin{equation*}
\alpha_{\tau_j}= \min\left\{1, \frac{\pi(\Theta^{*}|x) f_{\textnormal{N}}(\tau_{j}^{(s)}, V_{\tau_{j}})\mathbbm{1}_{(\tau_{j-1}^{(s+1)}, \tau_{j+1}^{(s)}) }
}{\pi(\tilde{\Theta}|x) f_{\textnormal{N}}(\tau_{j}^{*}, V_{\tau_{j}})\mathbbm{1}_{(\tau_{j-1}^{(s+1)}, \tau_{j+1}^{(s)})}}\right\},
\end{equation*}
$\Theta^{*} = \{\Phi^{(s+1)}, \Psi^{(s+1)},\tau^{*}\}$, $\tau^{*} = \{ \tau_{<j}^{(s+1)}, \tau_{j}^{*}, \tau_{>j}^{(s)}\}$ and $\tilde{\Theta} = \{\Phi^{(s+1)}, \Psi^{(s+1)}, \tau_{<j}^{(s+1)},\tau_{\geq j}^{(s)}\}$. 
\\

\subsection{CMNPD}
The steps for the CMNPD are the same as for the CMGPD with the only difference that the parameters of the mixture of normals now need to be estimated, i.e. the means $\mu=\{\mu_1,\dots,\mu_l\}$ and the variances $\delta=\{\delta_1,\dots,\delta_l\}$. In this case $\Phi=\{\mu,\delta,p\}$. At each iteration $s$, the normal parameters are updated as follows: \\

\textbf{Sampling $\mu$:} The proposal kernel for $\mu_{z}$, $z\in[l]$, is taken as the Gamma distribution $ G(\mu_{z}^{(s)},{\mu_{z}^{(s)}}^ 2/V_{\mu_{z}}) \mathbbm{1}_{(\mu_{1}^{(s+1)} < \cdots < \mu_{z-1}^{(s+1)} < \mu_{z}^{(s)} < \cdots < \mu_{h}^{(s)})}
$
where  $V_{\mu_{z}}$ is chosen to ensure appropriate chain mixing.
So $\mu_{z}^{(s+1)}=\mu_{z}^{*}$ with probability $\alpha_{\mu_{z}}$, where
\begin{equation*}
\alpha_{\mu_{z}} = \min \left\{1, \frac{\pi(\Theta^{*}|x) f_{\textnormal{G}}(\mu_{z}^{(s)}|\mu_{z}^{*},{\mu_{z}^{*}}^2 /V_{\mu_{z}})\mathbbm{1}_{ (\mu_{1}^{(s+1)} < \cdots < \mu_{z}^{*} < \cdots < \mu_{h}^{(s)})}
}{\pi(\tilde{\Theta}|x) f_{\textnormal{G}}(\mu_{z}^{*}|\mu_{z}^{(s)},{\mu_{z}^{(s)}}^ 2/V_{\mu_{z}}) \mathbbm{1}_{ (\mu_{1}^{(s+1)} < \cdots < \mu_{z}^{(s)} < \cdots < \mu_{h}^{(s)})}}\right\},
\end{equation*}
$\Theta^{*} = \{\mu^{*}, \delta^{(s)}, p^{(s)}, \Psi^{(s+1)},\tau^{(s)}\}$, $\mu^{*} = \{ \mu_{<z}^{(s+1)}, \mu_{z}^{*}, \mu_{>z}^{(s)}\}$ and $\tilde{\Theta} = \{\mu_{< z}^{(s+1)},\mu_{\geq z}^{(s)}, \delta^{(s)}, p^{(s)}, \Psi^{(s+1)},\tau^{(s)}\}$. 
\\

\textbf{Sampling $\delta$: } The proposal kernel for $\delta_{z}$, $z\in[l]$,  is taken as the Gamma distribution
$
G(\delta_{z}^{(s)},{\delta_{z}^{(s)}}^ 2/V_{\delta_{z}}) 
$
where $V_{\delta_{z}}$ is  chosen to ensure appropriate chain mixing.
So $\delta_{z}^{(s+1)}=\delta_{z}^{*}$  with probability $\alpha_{\delta_{z}}$, where
\begin{equation*}
\alpha_{\delta_{z}} = \min \left\{1, \frac{\pi(\Theta^{*}|x) f_{\textnormal{G}}(\delta_{z}^{(s)}|\delta_{z}^{*},{\delta_{z}^{*}}^2 /V_{\delta_{z}})
}{\pi(\tilde{\Theta}|x) f_{\textnormal{G}}(\delta_{z}^{*}|\delta_{z}^{(s)},{\delta_{z}^{(s)}}^ 2/V_{\delta_{z}})}\right\},
\end{equation*}
$\Theta^{*} = \{\mu^{(s+1)}, \delta^{*}, p^{(s)}, \Psi^{(s+1)},\tau^{(s)}\}$,  $\delta^{*} = \{ \delta_{<z}^{(s+1)}, \delta_{z}^{*}, \delta_{>z}^{(s)}\}$ and $\tilde{\Theta} = \{\mu^{(s+1)}, \delta_{< z}^{(s+1)},\delta_{\geq z}^{(s)}, p^{(s)}, \Psi^{(s+1)},\tau^{(s)}\}$. 
\end{document}